\documentclass[aps,prd,reprint,superscriptaddress,nofootinbib]{revtex4-2}
\usepackage{graphicx}
\usepackage{dcolumn}
\usepackage{bm}
\usepackage{amssymb}
\usepackage{amsmath}
\usepackage{amsfonts}
\usepackage{mathrsfs}
\usepackage{xcolor}
\usepackage{mathtools}

\usepackage[
	colorlinks=true,
	linkcolor=blue,
	citecolor=blue,
	urlcolor=blue
]{hyperref}

		\newcommand{\bea}{\begin{eqnarray*}} 
		\newcommand{\eea}{\end{eqnarray*}}
		\newcommand{\beq}{\begin{equation*}} 
		\newcommand{\eeq}{\end{equation*}}

\newcommand{\qed}{\nobreak \ifvmode \relax \else
      \ifdim\lastskip<1.5em \hskip-\lastskip
      \hskip1.5em plus0em minus0.5em \fi \nobreak
      \vrule height0.75em width0.5em depth0.25em\fi}

\begin{document}


\title{Quantum vacuum interaction between semitransparent plates in a
Pöschl--Teller background: an exact relative $TGTG$ formula}

\author{Jose M. Muñoz-Castañeda}	\email{jose.munoz.castaneda@uva.es}
\affiliation{Departamento de F\'isica Te\'orica, At\'omica y Optica,Universidad de Valladolid,
  Valladolid, 47011, Spain.} 
  \affiliation{IMUVa,Universidad de Valladolid,
  Valladolid, 47011, Spain.} 

\author{In\'es Cavero-Pel\'aez}
\affiliation{Departamento de F\'isica Te\'orica,
	Universidad de Zaragoza, 50009 Zaragoza, Spain}
	\affiliation{Centro de Astropartículas y Física de Altas Energías, Universidad de Zaragoza, 50009 Zaragoza, Spain}

\author{Gonzalo Sancho-Garrido}	
\affiliation{Departamento de F\'isica Te\'orica, At\'omica y Optica,Universidad de Valladolid,
  Valladolid, 47011, Spain.}

\begin{abstract}
We derive an exact background-dressed $TGTG$ formula for the one-loop
quantum vacuum interaction between two semitransparent Dirac-delta
plates in the one-soliton P\"oschl--Teller background. The interaction
is defined through a relative determinant in which the two isolated
one-plate contributions are subtracted in the same background. As a
result, the energy depends on the individual plate positions as well as
on their separation. For repulsive plates, the normalizable zero mode
does not induce an infrared divergence in the relative interaction
energy. We also extend the construction to planar plates and analyse the
corresponding position-dependent interaction forces.
\end{abstract}


\maketitle

\section{Introduction}
\label{sec:introduction}

The Casimir effect is one of the standard manifestations of the
dependence of quantum vacuum fluctuations on external conditions. In
its original form it describes the interaction induced by the
electromagnetic vacuum between two perfectly conducting plates
\cite{casimir1948}. More generally, the same mechanism appears in
quantum field theories in the presence of boundaries, interfaces,
background potentials, or nontrivial material response
\cite{milton2001book,bordag2009book}. From this point of view,
boundary conditions and singular interactions provide effective
descriptions of the way in which macroscopic or mesoscopic objects
modify the spectrum of quantum fluctuations.

A useful idealization of semitransparent plates is given by
point-supported interactions. In one spatial dimension, Dirac
$\delta$ potentials provide the simplest model of partially transparent
mirrors. They interpolate between weakly coupled transparent defects
and ideal Dirichlet boundaries in the strong-coupling limit. The
corresponding quantum vacuum interaction between two $\delta$ plates
has been studied by several methods, including Green functions,
stress-tensor techniques, phase shifts, and the $TGTG$ formalism
\cite{bordag1992delta,milton2004casimir,parashar2012delta,munoz2013prd}. The same strategy can be extended
to more general point interactions, such as $\delta$-$\delta'$ plates,
where generalized Robin-type boundary conditions and both attractive
and repulsive regimes arise \cite{munoz2015ddp}. These examples show
that point interactions are well adapted to analytically tractable
Casimir systems, provided that their spectral and self-adjointness
properties are kept under control.

In this paper we consider a different, but physically natural,
situation. The plates are not placed in a translation-invariant vacuum.
Instead, they are embedded in a smooth wall-like background. Such
backgrounds occur in field theories with degenerate vacua, where kink
or domain-wall configurations interpolate between different asymptotic
phases. Domain walls are standard extended topological defects in
cosmology and field theory \cite{vilenkinShellard1994}. They also
appear in gravitational contexts, where planar walls lead to
nontrivial space-time geometries and characteristic gravitational
effects \cite{ipserSikivie1984}. At the level of linearized scalar
fluctuations, the one-dimensional kink problem gives rise to
Schrödinger operators with reflectionless Pöschl--Teller potentials,
a standard structure in the semiclassical quantization of solitons
\cite{rajaraman1982,DHN1974}.

This observation motivates the model studied below. We regard the
Pöschl--Teller potential as an analytically controllable scalar
background capturing the one-dimensional transverse fluctuation
problem associated with a wall-like object. The plates are then
localized defects placed in the presence of this background. The
resulting system is a simple model for quantum vacuum forces between
semitransparent probes in the neighbourhood of an extended object. It
also provides a tractable setting in which to study how the usual
Casimir interaction is modified when translation invariance is broken
at the level of the fixed-background quadratic fluctuation theory.

The reference fluctuation operator is the one-soliton Pöschl--Teller
operator
\begin{equation*}
	K_{\rm PT}
	=
	-\frac{d^2}{dx^2}
	+
	1
	-
	2\,{\rm sech}^2x ,
	\label{eq:intro_KPT}
\end{equation*}
in dimensionless units. Its continuous spectrum starts at $1$, and it
has a normalizable zero mode. This operator also has a direct solitonic
interpretation: it is the small-fluctuation operator around the
sine-Gordon kink $\phi_K(x)=4\arctan e^x$, since
$V''(\phi_K)=\cos\phi_K=1-2\operatorname{sech}^2x$. Its zero mode is
proportional to $\phi'_K(x)$ and therefore represents the translational
Goldstone mode of the kink. This identification provides a concrete
physical realization of the model: two semitransparent probes embedded
in the fluctuation background of a topological kink, or, in the planar
extension, plates parallel to a domain-wall profile. Although the
underlying sine-Gordon theory is translationally invariant, here the
kink is chosen to be centred at $x=0$ and the corresponding
Pöschl--Teller background is kept fixed. The resulting quadratic
fluctuation problem is therefore not translationally invariant. The
corresponding normalizable zero mode encodes the translational symmetry
of the underlying theory and is fully retained in the fluctuation
determinant; what is not included is a dynamical treatment of the kink
position beyond the quadratic approximation.

For the purposes of the present construction, the Pöschl--Teller
background plays a second role. Besides providing a concrete kink
realization, it is an exactly solvable prototype in which the
background-dressed relative determinant can be written explicitly. We
do not assume, nor prove here, a general formula for arbitrary
inhomogeneous backgrounds. The point is rather that this solvable
background allows the relative-determinant construction to be exhibited
with full spectral control. In particular, the interaction between
two plates can no longer depend only on their separation. It depends on
their individual positions with respect to the centre of the
Pöschl--Teller background.

We introduce two semitransparent plates by adding Dirac interactions
at independent positions $a_1<a_2$,
\begin{equation*}
	V_\delta(x)
	=
	\lambda_1\delta(x-a_1)
	+
	\lambda_2\delta(x-a_2).
	\label{eq:intro_delta_plates}
\end{equation*}
The full fluctuation operator is therefore
\begin{equation*}
	K_{\lambda_1,\lambda_2}
	=
	K_{\rm PT}
	+
	\lambda_1\delta(x-a_1)
	+
	\lambda_2\delta(x-a_2).
	\label{eq:intro_full_operator}
\end{equation*}
Before computing any vacuum energy, one must identify the coupling
domain in which the fluctuation problem defines a stable quantum field
theory. This point is essential in the present background. Since
$K_{\rm PT}$ has a normalizable zero mode, an isolated attractive delta
plate destabilizes the unshifted P\"oschl--Teller background. By
contrast, for
\begin{equation*}
	\lambda_1\geq0,
	\qquad
	\lambda_2\geq0,
	\label{eq:intro_stable_domain}
\end{equation*}
the repulsive interactions cannot generate negative normal modes. This
therefore provides a manifestly stable domain for the relative
construction, including the two isolated one-plate configurations that
enter the subtraction. It is not intended to characterize the full
stability domain of the combined two-plate problem. To include
attractive plates within the relative construction, the reference
problem must be modified. A positive mass shift provides one such
modification, with the admissible coupling domain determined by the
corresponding Birman--Schwinger conditions
\cite{ReedSimonIV1978,SimonTraceIdeals2005}.

The main goal of the paper is to derive a new exact background-dressed
$TGTG$ formula for the interaction part of the vacuum energy of two
semitransparent plates. The formula follows from a relative-determinant
construction in which the subtraction prescription is essential. We subtract
the two isolated one-plate contributions in the same P\"oschl--Teller
background. Hence the result is not obtained by a formal replacement of the
free Green function by the P\"oschl--Teller Green function in the
translation-invariant double-delta formula. Rather, it is a relative
determinant with respect to the background operator, with all one-plate
subtractions performed in that same background.

Let $G_\xi^{\rm PT}$ denote the Euclidean Green function of
$K_{\rm PT}$ and set
\begin{equation*}
	G_{ij}(\xi)
	=
	G_\xi^{\rm PT}(a_i,a_j),
	\qquad
	i,j=1,2.
	\label{eq:intro_Gij}
\end{equation*}
The interaction energy can be written as
\begin{equation*}
	E_{\rm int}
	=
	\frac{1}{2\pi}
	\int_0^\infty d\xi\,
	\log\Delta(\xi),
	\label{eq:intro_Eint_compact}
\end{equation*}
where $\Delta(\xi)$ is the relative determinant obtained by dividing
the two-plate determinant by the two one-plate determinants. The
finite-rank reduction gives
\begin{equation*}
	\Delta(\xi)
	=
	1-
	\frac{
		\lambda_1\lambda_2G_{12}(\xi)^2
	}{
		\left(1+\lambda_1G_{11}(\xi)\right)
		\left(1+\lambda_2G_{22}(\xi)\right)
	}.
	\label{eq:intro_delta_xi}
\end{equation*}
The pole of $G_\xi^{\rm PT}$ at $\xi=0$, caused by the zero mode, does
not invalidate the interaction formula. It produces only an integrable
logarithmic singularity in the improper integral. At large plate
separation, the interaction is controlled by the infrared scale
associated with the zero-mode pole. On this scale, the full relative
determinant approaches the same zero-mode relative-determinant structure
as the pure zero-mode projection, but with the coupling of each plate
dressed by the local response of the massive fluctuations. Their direct
propagation between widely separated plates is subleading.

The $TGTG$ formula gives a compact scattering-theoretic expression for
the interaction part of the vacuum energy between two disjoint
objects. In the formulation of Kenneth and Klich, the interaction is
expressed in terms of transition operators and the Green function of
the reference problem \cite{KennethKlich2008}. In the present
one-dimensional setting the same structure survives, but the reference
problem is the Pöschl--Teller operator rather than the free operator.
Equivalently, the result may be expressed in terms of left and right
reflection data for a delta defect in the Pöschl--Teller background.
This scattering interpretation is useful, but the determinant
derivation fixes the subtraction prescription unambiguously.

Related transfer-operator constructions for sine-Gordon-type
backgrounds have been considered elsewhere
\cite{casimirTransferSG}. The present treatment differs in three
essential points. First, the interaction energy is derived directly
from the relative determinant of the Pöschl--Teller resolvent.
Second, the one-plate contributions are subtracted in the same
background. Third, the positions $a_1$ and $a_2$ are kept independent
throughout. This is necessary because the quadratic action governing
the fluctuations about the fixed background is not translationally
invariant.

The resulting energy
\[
E_{\rm int}
=
E_{\rm int}(a_1,a_2;\lambda_1,\lambda_2)
\]
contains effects absent from the free double-delta problem. Because it
depends separately on $a_1$ and $a_2$, the relative interaction energy
responds both to a change in the plate separation and to a rigid
displacement of the pair with respect to the fixed Pöschl--Teller
profile. The latter is an interaction contribution associated with the
rigid displacement of the pair, not the total mechanical force on the
two-plate system, which would also contain the position dependence of
the individual one-plate contributions. This position dependence
reflects the fact that the quadratic action governing the fluctuations
about the fixed background is not translationally invariant and is not
in conflict with momentum conservation in the underlying
translationally invariant theory. These position-dependent effects are
the first direct physical consequence of the background-dressed
relative determinant.

We stress that the present work is restricted to the interaction part of
the vacuum energy. The one-plate contributions subtracted in the
relative determinant are genuine vacuum energies of isolated plates in
the same Pöschl--Teller background and depend on the plate position.
They would therefore contribute separately to the total mechanical
force. Their analysis, including the corresponding renormalization
problem, constitutes a distinct one-loop problem and is the subject of
ongoing work.

The paper is organized as follows. In
Sec.~\ref{sec:scalar_fluctuations} we define the scalar fluctuation
problem, fix the dimensionless Pöschl--Teller operator, and introduce
the two delta plates. In Sec.~\ref{sec:bound_states_stability} we
identify the manifestly stable coupling domain of the unshifted
fluctuation problem. In Sec.~\ref{sec:PT_green_function} we construct
the Euclidean Green function describing propagation through the
Pöschl--Teller background. Section~\ref{sec:finite_rank_plates} reduces
the two-plate problem to a finite-dimensional Birman--Schwinger
determinant, and Sec.~\ref{sec:relative_determinant_TGTG} derives the
relative determinant and the corresponding background-dressed $TGTG$
formula, including its zero-mode and scattering descriptions.
Limiting regimes and the controlled large-separation asymptotics are
analysed in Sec.~\ref{sec:limiting_regimes}. In
Sec.~\ref{sec:position_dependent_interaction} we extend the construction
to planar plates, analyse the position dependence of the interaction,
and study the corresponding interaction-force densities. Technical
details on the Green function, finite-rank determinants, infrared
behaviour, Birman--Schwinger stability and one-dimensional scattering
are collected in the appendices.

\section{Scalar fluctuations, Pöschl--Teller background and delta plates}
\label{sec:scalar_fluctuations}

We consider one real scalar field in $(1+1)$-dimensional Minkowski
space. Natural units $\hbar=c=1$ are used throughout. Dimensionful
variables will be denoted with tildes only in this paragraph. The quadratic action for fluctuations in a fixed static background is
\begin{equation*}
	\widetilde S[\Phi]
	=
	\frac12
	\int_{\mathbb R^2}d\widetilde t\,d\widetilde x\,
	\left[
	(\partial_{\widetilde t}\Phi)^2
	-
	(\partial_{\widetilde x}\Phi)^2
	-
	\widetilde U(\widetilde x)\Phi^2
	\right],
	\label{eq:dimensionful_action}
\end{equation*}
where the background is asymptotically massive,
\begin{equation*}
	\lim_{|\widetilde x|\to\infty}
	\widetilde U(\widetilde x)=m^2,
	\qquad
	m>0 .
	\label{eq:dimensionful_asymptotic_mass}
\end{equation*}
We introduce dimensionless variables by
\begin{equation*}
	t=m\widetilde t,
	\qquad
	x=m\widetilde x,
	\qquad
	\omega=\frac{\widetilde\omega}{m},
	\qquad
	\widetilde U(\widetilde x)=m^2U(x).
	\label{eq:dimensionless_variables}
\end{equation*}
Since the scalar field is dimensionless in two space-time dimensions, no
field rescaling is required, so that $\phi(t,x)=\Phi(\tilde t,\tilde x)$
in the rescaled coordinates. From now on all quantities are dimensionless
unless explicitly stated otherwise.

The action becomes
\begin{equation*}
	S[\phi]
	=
	\frac12
	\int_{\mathbb R^2}dt\,dx\,
	\left[
	(\partial_t\phi)^2
	-
	(\partial_x\phi)^2
	-
	U(x)\phi^2
	\right],
	\label{eq:dimensionless_action}
\end{equation*}
and the normal modes of the quantum fluctuation field are characterized
by the spectral problem
\begin{equation*}
	K\psi=\omega^2\psi,
	\qquad
	K=-\frac{d^2}{dx^2}+U(x).
	\label{eq:dimensionless_spectral_problem}
\end{equation*}

The smooth background used in this work is the one-soliton
Pöschl--Teller potential
\begin{equation*}
	U_{\rm PT}(x)=1-2\,{\rm sech}^2x .
	\label{eq:PT_potential}
\end{equation*}
Accordingly,
\begin{equation}
	K_{\rm PT}
	=
	-\frac{d^2}{dx^2}
	+
	1
	-
	2\,{\rm sech}^2x .
	\label{eq:KPT_def}
\end{equation}
The operator $K_{\rm PT}$ is self-adjoint on
$H^2(\mathbb R)$,\footnote{Here $H^n(\mathbb R)$ denotes the Sobolev
space of square-integrable functions whose weak derivatives up to order
$n$ are also square integrable. In particular, $H^2(\mathbb R)$ is the
operator domain of $K_{\rm PT}$, while $H^1(\mathbb R)$ is its quadratic-form
domain.}
has essential spectrum
\begin{equation*}
	\sigma_{\rm ess}(K_{\rm PT})=[1,\infty),
	\label{eq:KPT_essential_spectrum}
\end{equation*}
and possesses a single normalized bound state at zero eigenvalue,
\begin{equation}
	\psi_0(x)=\frac{1}{\sqrt2}\,{\rm sech}\,x,
	\qquad
	K_{\rm PT}\psi_0=0 .
	\label{eq:PT_zero_mode}
\end{equation}
The zero mode will play an important role in the stability analysis.

Two semitransparent plates are introduced by adding Dirac delta
interactions at independent positions $a_1<a_2$. In dimensionless
form,
\begin{equation*}
	V_\delta(x)
	=
	\lambda_1\delta(x-a_1)
	+
	\lambda_2\delta(x-a_2),
	\label{eq:delta_potential_dimensionless}
\end{equation*}
and the full fluctuation operator is
\begin{equation}
	K_{\lambda_1,\lambda_2}
	=
	K_{\rm PT}
	+
	\lambda_1\delta(x-a_1)
	+
	\lambda_2\delta(x-a_2).
	\label{eq:full_fluctuation_operator}
\end{equation}
If the dimensionful couplings are denoted by $\widetilde\lambda_i$,
then
\begin{equation*}
	\lambda_i=\frac{\widetilde\lambda_i}{m},
	\qquad
	a_i=m\widetilde a_i .
	\label{eq:dimensionless_delta_couplings}
\end{equation*}

The delta interactions are understood through local matching
conditions at their support, in the standard point-interaction
description of one-dimensional Schrödinger operators
\cite{albeverio2005,barton1993waxman}. Equivalently, the boundary data
on the two sides of each plate are related by
\begin{equation}
	\begin{pmatrix}
		\psi(a_i^+)\\[1mm]
		\psi'(a_i^+)
	\end{pmatrix}
	=
	\begin{pmatrix}
		1 & 0\\[1mm]
		\lambda_i & 1
	\end{pmatrix}
	\begin{pmatrix}
		\psi(a_i^-)\\[1mm]
		\psi'(a_i^-)
	\end{pmatrix},
	\qquad
	i=1,2 .
	\label{eq:delta_matching_matrix}
\end{equation}
The delta interaction is a quadratic-form perturbation on
$H^1(\mathbb R)$, rather than a bounded finite-rank operator on
$L^2(\mathbb R)$. Let
\begin{equation*}
	\Gamma:H^1(\mathbb R)\longrightarrow\mathbb C^2,
	\qquad
	\Gamma\psi
	=
	\begin{pmatrix}
		\psi(a_1)\\
		\psi(a_2)
	\end{pmatrix},
\end{equation*}
be the trace map at the two plate positions. The delta contribution to
the quadratic form is
\begin{equation*}
	q_\delta[\psi]
	=
	\lambda_1|\psi(a_1)|^2
	+
	\lambda_2|\psi(a_2)|^2 .
\end{equation*}
In Dirac notation this form perturbation may be represented formally as
\begin{equation}
	V_\delta
	=
	\lambda_1|a_1\rangle\langle a_1|
	+
	\lambda_2|a_2\rangle\langle a_2| .
	\label{eq:delta_finite_rank_operator}
\end{equation}
The finite-rank reduction used below arises only after the trace map is
composed with the background resolvent.

\section{Stability of the fluctuation problem}
\label{sec:bound_states_stability}

Before computing the vacuum interaction energy, one must specify the
coupling domain in which the fluctuation operator defines a stable
quantum field theory. In the present model this is not a purely
technical point. The Pöschl--Teller operator introduced in Sec.~II has a
normalizable zero mode, so an attractive point perturbation can move
this mode to negative $\omega^2$. For real couplings, the spatial
operator defined by the matching conditions remains self-adjoint.
However, the quantum field theory defined by the quadratic action for
fluctuations requires in addition that this operator be non-negative.
Indeed, the corresponding field Hamiltonian is constructed as a sum
of harmonic-oscillator Hamiltonians with normal-mode frequencies
$\omega$. If $\omega^2<0$, the corresponding frequency becomes
imaginary and this construction no longer yields a self-adjoint field
Hamiltonian with a stable vacuum \cite{MunozCastaneda2013}. We therefore restrict the
fluctuation QFT considered here to the sector in which
$K_{\lambda_1,\lambda_2}$ is non-negative.

Let $q_{\rm PT}$ denote the quadratic form associated with
$K_{\rm PT}$. The quadratic form of the two-plate operator defined in
Eq.~\eqref{eq:full_fluctuation_operator} is, for
$\psi\in H^1(\mathbb R)$,
\begin{align*}
	\mathcal Q_{\lambda_1,\lambda_2}[\psi]
	&=
	q_{\rm PT}[\psi]
	+
	\lambda_1|\psi(a_1)|^2
	+
	\lambda_2|\psi(a_2)|^2 .
\end{align*}
Since $K_{\rm PT}\geq0$, it follows immediately that
\begin{equation*}
	\lambda_1\geq0,
	\,\,
	\lambda_2\geq0
	\,\,
	\Longrightarrow
	\,\,
	\mathcal Q_{\lambda_1,\lambda_2}[\psi]\geq0
	\,\,
	\forall \,\,
	\psi\in H^1(\mathbb R).
	\label{eq:stable_domain_nonnegative_form}
\end{equation*}
Therefore
\begin{equation*}
	\lambda_1\geq0,
	\qquad
	\lambda_2\geq0
	\quad
	\Longrightarrow
	\quad
	K_{\lambda_1,\lambda_2}\geq0 .
	\label{eq:stable_domain_operator}
\end{equation*}
This proves absence of negative normal modes in the repulsive sector.

For attractive couplings, the preceding quadratic-form argument no
longer establishes non-negativity. One must then determine whether the
point perturbations produce a negative eigenvalue. This can be done by
the Birman--Schwinger principle
\cite{ReedSimonIV1978,SimonTraceIdeals2005}: instead of solving the
eigenvalue problem for the perturbed operator directly, one rewrites
the negative-mode condition in terms of the resolvent of the
unperturbed Pöschl--Teller operator evaluated at the support of the
point perturbations. For one delta this gives a scalar equation; for
two deltas it gives a finite-dimensional determinant condition. The
details are recalled in Appendix~\ref{app:birman_schwinger}.

The physical consequence is simple. In the unshifted
model, an isolated attractive delta already destabilizes
the zero mode of the P\"oschl--Teller background. Indeed,
using the zero mode \eqref{eq:PT_zero_mode} as a trial function in the
one-plate quadratic form gives
\[
q_{\lambda,a}[\psi_0]
=
q_{\rm PT}[\psi_0]+\lambda|\psi_0(a)|^2
=
\lambda|\psi_0(a)|^2 .
\]
Since $\psi_0(a)\neq0$ for every finite $a$, this quantity is
negative for every $\lambda<0$. Hence an attractive plate
produces a mode with $\omega^2<0$ in the unshifted theory.

In the rest of the paper we work in the manifestly stable sector
\begin{equation*}
	\lambda_1\geq0,
	\qquad
	\lambda_2\geq0 .
	\label{eq:stable_sector_assumption}
\end{equation*}
In this domain the two one-plate operators and the full two-plate
operator entering the relative determinant are all non-negative.
Since an isolated attractive plate destabilizes the zero mode in the
unshifted background, stable attractive configurations require a
modification of the reference operator. A positive mass shift provides
such a modification, subject to the corresponding Birman--Schwinger
stability conditions discussed in Appendix~\ref{app:birman_schwinger}.

\section{Background-dressed propagation and the Pöschl--Teller Green function}
\label{sec:PT_green_function}

The next ingredient is the Euclidean propagator of the fluctuation
field in the wall-like background. In a translation-invariant vacuum,
the Green function depends only on the relative coordinate $x-y$, and
all information about propagation between two plates is reduced to
their separation. This is no longer true in the present problem. The
P\"oschl--Teller background selects a distinguished point, the centre of
the domain-wall profile, so that the quadratic action governing
fluctuations around the fixed background is not translationally invariant.
Consequently, the propagation of a fluctuation from $y$ to $x$ depends
on both positions separately. The Green function of the background
operator is precisely the object that encodes this dressed propagation.
It will enter the relative determinant through its values at the plate
positions and is ultimately responsible for the dependence of the
interaction energy on $a_1$ and $a_2$, not only on $a_2-a_1$.

For $\xi>0$ we define the Euclidean resolvent of the background
operator by
\begin{equation*}
	G_\xi^{\rm PT}
	=
	\left(K_{\rm PT}+\xi^2\right)^{-1},
	\qquad
	\xi>0,
	\label{eq:secIV_reference_resolvent}
\end{equation*}
with kernel satisfying
\begin{equation*}
	\left(K_{\rm PT}+\xi^2\right)G_\xi^{\rm PT}(x,y)
	=
	\delta(x-y).
	\label{eq:secIV_green_equation}
\end{equation*}
The endpoint $\xi=0$ is singular because the background operator has
the zero mode given in Eq.~\eqref{eq:PT_zero_mode}.

Set
\begin{equation}
	\kappa=\sqrt{1+\xi^2},
	\label{eq:kappa_definition}
\end{equation}
and let $x_< =\min\{x,y\}$, $x_> =\max\{x,y\}$. The Pöschl--Teller
Green function is
\begin{equation}
	G_\xi^{\rm PT}(x,y)
	=
	\frac{
		e^{-\kappa|x-y|}
		(\kappa-\tanh x_<)(\kappa+\tanh x_>)
	}{
		2\kappa(\kappa^2-1)
	} .
	\label{eq:PT_green_explicit_xi}
\end{equation}
Since Eq.~\eqref{eq:kappa_definition} gives
$\kappa^2-1=\xi^2$, the denominator in
Eq.~\eqref{eq:PT_green_explicit_xi} already displays the infrared
$1/\xi^2$ singularity associated with the zero mode.
Its derivation from the scattering solutions of the reflectionless
background is given in App.~\ref{app:PT_green_function}. In the main
text we only need the consequences collected below.

First, reality and self-adjointness give the symmetry
\begin{equation*}
	G_\xi^{\rm PT}(x,y)=G_\xi^{\rm PT}(y,x).
	\label{eq:secIV_green_symmetry_background}
\end{equation*}
Second, for the plate positions $a_1<a_2$, we write
\begin{equation}
	G_{ij}(\xi)=G_\xi^{\rm PT}(a_i,a_j),
	\qquad
	i,j=1,2 .
	\label{eq:secIV_Gij_def}
\end{equation}
Using Eq.~\eqref{eq:PT_green_explicit_xi}, one obtains
\begin{align}
	G_{ii}(\xi)
	&=
	\frac{\xi^2+\operatorname{sech}^2 a_i}
	{2\kappa\xi^2},
	\qquad i=1,2,
	\label{eq:PT_green_diagonal_plates}
	\\
	G_{12}(\xi)
	&=
	\frac{
		e^{-\kappa(a_2-a_1)}
		(\kappa-\tanh a_1)(\kappa+\tanh a_2)
	}{2\kappa\xi^2} .
	\label{eq:PT_green_off_diagonal_plates}
\end{align}
These expressions already display the loss of translation invariance:
the diagonal entries depend on the individual positions of the plates,
and the off-diagonal entry is not a function of the separation alone.

The zero mode produces the infrared expansion
\begin{equation}
	G_\xi^{\rm PT}(x,y)
	=
	\frac{\psi_0(x)\psi_0(y)}{\xi^2}
	+
	O(1),
	\qquad
	\xi\to0^+ .
	\label{eq:PT_green_zero_mode_pole}
\end{equation}
The zero-mode pole is important, but in the repulsive sector considered
here it does not make the interaction energy infrared divergent. The cancellation in
the relative determinant is stated in
Sec.~\ref{sec:relative_determinant_TGTG} and proved in
App.~\ref{app:infrared_relative_determinant}.

Finally, for fixed plate separation $d=a_2-a_1>0$, the off-diagonal
entry has the large-frequency behaviour
\begin{equation}
	G_{12}(\xi)
	=
	O\!\left(\frac{e^{-\xi d}}{\xi}\right),
	\qquad
	\xi\to+\infty .
	\label{eq:PT_green_large_xi}
\end{equation}
This exponential decay will ensure ultraviolet convergence of the
interaction part of the vacuum energy.

\section{Finite-rank Birman--Schwinger reduction}
\label{sec:finite_rank_plates}

The two delta plates couple to the fluctuation field only through its
values at $x=a_1$ and $x=a_2$. For later use we write the individual
plate perturbations as
\begin{equation*}
    V_i(x)=\lambda_i\delta(x-a_i),
    \qquad i=1,2,
\end{equation*}
so that $V_\delta=V_1+V_2$. At fixed Euclidean frequency, the
two-plate problem is therefore controlled by the P\"oschl--Teller
Green function evaluated at the plate positions.\footnote{Strictly
speaking, the delta plates are quadratic-form perturbations rather than
bounded finite-rank operators on $L^2(\mathbb R)$. Their
finite-dimensional reduction follows from the trace map to the plate
positions. The associated symmetrized Birman--Schwinger operator is
trace class, and its Fredholm determinant reduces exactly to the
finite-dimensional determinant used below; see
App.~\ref{app:rank_two_determinants}.}

We collect these Green-function values in the matrix
\begin{equation}
    \mathbb G(\xi)
    =
    \begin{pmatrix}
        G_{11}(\xi) & G_{12}(\xi)
        \\
        G_{21}(\xi) & G_{22}(\xi)
    \end{pmatrix},
    \label{eq:secV_green_matrix}
\end{equation}
where $G_{ij}(\xi)=G_\xi^{\rm PT}(a_i,a_j)$ as defined in
Eq.~\eqref{eq:secIV_Gij_def}, with
$G_{12}(\xi)=G_{21}(\xi)$. The corresponding coupling matrix is
\begin{equation}
    \Lambda
    =
    \begin{pmatrix}
        \lambda_1 & 0
        \\
        0 & \lambda_2
    \end{pmatrix}.
    \label{eq:secV_lambda_matrix}
\end{equation}

For compactness, the common argument $\xi$ of the Green-function
entries will be suppressed in the following equations. The
Birman--Schwinger reduction gives the exact two-plate determinant
\begin{equation}
    \begin{aligned}
    D_{12}(\xi)
    &=
    \det_{\mathbb C^2}
    \left[
        \mathbb I_2+\Lambda\mathbb G(\xi)
    \right]
    \\
    &=
    \det
    \begin{pmatrix}
        1+\lambda_1G_{11} & \lambda_1G_{12}
        \\
        \lambda_2G_{21} & 1+\lambda_2G_{22}
    \end{pmatrix}.
    \end{aligned}
    \label{eq:secV_two_plate_det_matrix}
\end{equation}
Hence
\begin{equation}
    \begin{aligned}
    D_{12}(\xi)
    &=
    (1+\lambda_1G_{11})
    (1+\lambda_2G_{22})
    \\
    &\quad
    -\lambda_1\lambda_2G_{12}G_{21}.
    \end{aligned}
    \label{eq:secV_two_plate_det}
\end{equation}
The first term contains the local response of the two isolated plates
in the P\"oschl--Teller background, whereas the second term couples the
two plate positions through propagation between them. Thus, for the
purpose of the interaction determinant, the original fluctuation
problem reduces exactly to a $2\times2$ problem involving only
propagation between, and back to, the plate positions.

For a single plate the same reduction gives
\begin{equation}
    D_i(\xi)
    =
    1+\lambda_iG_{ii}(\xi),
    \qquad i=1,2.
    \label{eq:secV_single_plate_det}
\end{equation}
Crucially, $D_i$ describes an isolated plate embedded in the same
P\"oschl--Teller background, rather than a free-space auxiliary
problem. The interaction contribution will therefore be obtained by
dividing $D_{12}$ by $D_1D_2$, thereby removing the two one-plate
contributions in the same reference background.

\section{Relative determinant and background-dressed $TGTG$ formula}
\label{sec:relative_determinant_TGTG}

We now turn the determinant reduction of Sec.~\ref{sec:finite_rank_plates}
into the exact interaction energy. All determinants are defined relative
to the same P\"oschl--Teller fluctuation operator $K_{\rm PT}$. Thus the
one-body subtraction must also be performed in that same background.

Let $E_{12}$ denote the one-loop vacuum-energy shift produced by the
two plates and $E_i$ that produced by plate $i$ alone, always with
$K_{\rm PT}$ as the reference operator. The interaction energy is
defined by subtracting the two isolated one-plate contributions,
\begin{equation}
    E_{\rm int}
    =
    E_{12}-E_1-E_2 .
    \label{eq:interaction_energy_subtraction}
\end{equation}
Individually, these vacuum-energy shifts contain the one-plate
ultraviolet self-energies and require regularization. With the same
regulator understood in all three terms, their Birman--Schwinger
representations are
\begin{equation}
    E_{12}
    =
    \frac{1}{2\pi}
    \int_0^\infty d\xi\,
    \log D_{12}(\xi),
\end{equation}
whereas for each isolated plate
\begin{equation}
    E_i
    =
    \frac{1}{2\pi}
    \int_0^\infty d\xi\,
    \log D_i(\xi),
    \qquad i=1,2 .
\end{equation}
All three determinants are defined with respect to the same
P\"oschl--Teller reference operator. Substituting these expressions
into Eq.~\eqref{eq:interaction_energy_subtraction} gives
\begin{equation}
    E_{\rm int}
    =
    \frac{1}{2\pi}
    \int_0^\infty d\xi\,
    \left[
        \log D_{12}(\xi)
        -
        \log D_1(\xi)
        -
        \log D_2(\xi)
    \right].
\end{equation}
The ultraviolet-sensitive one-plate contributions cancel in this
relative combination. The subtraction can therefore be performed
frequency by frequency, which naturally defines the relative
determinant
\begin{equation}
    \Delta(\xi)
    =
    \frac{D_{12}(\xi)}
    {D_1(\xi)D_2(\xi)} .
    \label{eq:secV_Delta_def}
\end{equation}
Hence
\begin{equation}
    E_{\rm int}
    =
    \frac{1}{2\pi}
    \int_0^\infty d\xi\,
    \log\Delta(\xi).
    \label{eq:interaction_energy_explicit_relative_det}
\end{equation}

The content of the relative determinant becomes explicit by using the
results of Sec.~\ref{sec:finite_rank_plates}. From
Eqs.~\eqref{eq:secV_two_plate_det} and
\eqref{eq:secV_single_plate_det},
\begin{equation}
    \Delta(\xi)
    =
    1-
    \frac{
        \lambda_1\lambda_2
        G_{12}(\xi)G_{21}(\xi)
    }{
        D_1(\xi)D_2(\xi)
    } .
    \label{eq:relative_det_explicit_D}
\end{equation}
Thus the product $D_1D_2$, which contains the two isolated one-plate
responses, divides out from the disconnected part of $D_{12}$.
Its remaining appearance in the denominator dresses the term that
connects the two plate positions.

This structure is most naturally expressed through the scalar
one-plate transition amplitudes
\begin{equation}
    \tau_i(\xi)
    =
    \frac{\lambda_i}{D_i(\xi)}
    =
    \frac{\lambda_i}
    {1+\lambda_iG_{ii}(\xi)},
    \qquad i=1,2 .
    \label{eq:single_plate_tau}
\end{equation}
Each $\tau_i$ resums repeated propagation from plate $i$ back to the
same plate through the P\"oschl--Teller background. Equation
\eqref{eq:relative_det_explicit_D} therefore becomes
\begin{equation}
    \Delta(\xi)
    =
    1-
    \tau_1(\xi)G_{12}(\xi)
    \tau_2(\xi)G_{21}(\xi).
    \label{eq:relative_determinant_TGTG_scalar_form}
\end{equation}
The term subtracted from unity has a direct multiple-scattering
interpretation: scattering at the first plate, propagation to the
second, scattering there, and propagation back to the first. It is
therefore convenient to collect this complete two-plate scattering
process into the factor
\begin{equation}
    \mathcal R(\xi)
    =
    \tau_1(\xi)G_{12}(\xi)
    \tau_2(\xi)G_{21}(\xi),
    \label{eq:secV_R_def}
\end{equation}
so that
\begin{equation}
    \Delta(\xi)
    =
    1-\mathcal R(\xi).
    \label{eq:secV_Delta_R}
\end{equation}
Using $G_{12}=G_{21}$, this factor can be written explicitly as
\begin{equation}
    \mathcal R(\xi)
    =
    \frac{
        \lambda_1\lambda_2G_{12}(\xi)^2
    }{
        [1+\lambda_1G_{11}(\xi)]
        [1+\lambda_2G_{22}(\xi)]
    } .
    \label{eq:secV_R_explicit}
\end{equation}

The operator $TGTG$ structure underlying
Eq.~\eqref{eq:relative_determinant_TGTG_scalar_form} is obtained by
introducing the one-plate transition operators with respect to the same
P\"oschl--Teller reference propagator,
\begin{equation}
    T_i(\xi)
    =
    V_i
    \left(
        1+G_\xi^{\rm PT}V_i
    \right)^{-1},
    \qquad i=1,2 .
    \label{eq:single_plate_T_operators}
\end{equation}
For a point-supported plate, $T_i$ is completely characterized by the
scalar amplitude $\tau_i(\xi)$.\footnote{In the customary
distributional shorthand one may write
$T_i(\xi)=\tau_i(\xi)|a_i\rangle\langle a_i|$, corresponding to the
kernel
$T_i(\xi;x,y)=\tau_i(\xi)\delta(x-a_i)\delta(y-a_i)$.
This representation makes the reduction to the one-dimensional
boundary space immediate.}

The exact relative determinant can therefore be written in the
background-dressed $TGTG$ form
\begin{equation}
    \Delta(\xi)
    =
    \det
    \left[
        1-
        T_1(\xi)G_\xi^{\rm PT}
        T_2(\xi)G_\xi^{\rm PT}
    \right].
    \label{eq:relative_determinant_TGTG_operator_form}
\end{equation}
For the present delta plates, its boundary-space reduction is precisely
Eq.~\eqref{eq:relative_determinant_TGTG_scalar_form}. Hence the relative
determinant obtained from the Birman--Schwinger reduction is exactly the
$TGTG$ determinant, with both the one-body transition operators and the
propagation between the plates defined with respect to the same
P\"oschl--Teller background.

Combining this result with
Eq.~\eqref{eq:interaction_energy_explicit_relative_det} gives
\begin{equation}
    \begin{aligned}
    E_{\rm int}
    &=
    \frac{1}{2\pi}
    \int_0^\infty d\xi\,
    \log\det\Big[
        1-T_1(\xi)G_\xi^{\rm PT}
    \\
    &\hspace{6.0em}
        {}\times T_2(\xi)G_\xi^{\rm PT}
    \Big].
    \end{aligned}
    \label{eq:secV_generalized_TGTG}
\end{equation}
The corresponding dimensionful energy is $mE_{\rm int}$.

The qualifier ``background-dressed'' is essential. The off-diagonal
kernels $G_{12}$ and $G_{21}$ describe propagation between the plates
through the P\"oschl--Teller background, while the diagonal kernels
$G_{ii}$ enter the transition amplitudes $\tau_i$ and therefore dress
the local response of each individual plate. The result is thus not
obtained by merely replacing the free Green function by
$G_\xi^{\rm PT}$ in the free double-delta formula: propagation and
one-body response are consistently defined with respect to the same
nontrivial reference operator.

It remains to verify that the exact interaction energy is well defined
at the two integration endpoints. For this purpose it is useful to
separate the localised zero mode from the massive fluctuations. In the
notation used below, define
\begin{equation}
    \psi_i\equiv\psi_0(a_i),
    \qquad
    g^\perp_{ij}(\xi)
    \equiv
    G^\perp_\xi(a_i,a_j),
\end{equation}
where
\begin{equation}
    G_\xi^{\rm PT}(x,y)
    =
    \frac{\psi_0(x)\psi_0(y)}{\xi^2}
    +
    G^\perp_\xi(x,y).
    \label{eq:zero_mode_resolvent_decomposition}
\end{equation}
Accordingly,
\begin{equation}
    G_{ij}(\xi)
    =
    \frac{\psi_i\psi_j}{\xi^2}
    +
    g^\perp_{ij}(\xi).
    \label{eq:secVI_Gij_IR}
\end{equation}
The first term is the contribution of the localised zero-mode
fluctuations. The reduced Green function $G^\perp_\xi$ contains the
massive fluctuations and is regular at $\xi=0$, since the remainder of
the P\"oschl--Teller spectrum is separated from the zero mode by the
mass gap.

For repulsive plates, $\lambda_1,\lambda_2>0$, this separation inserted
in the exact relative determinant gives
\begin{equation}
    \Delta(\xi)
    =
    C_{\rm IR}\,\xi^2
    +
    O(\xi^4),
    \qquad
    C_{\rm IR}>0,
    \qquad
    \xi\to0^+ .
    \label{eq:secV_Delta_IR_summary}
\end{equation}
Thus the zero-mode pole does not generate an infrared divergence in the
interaction energy. Instead,
\begin{equation}
    \log\Delta(\xi)
    =
    2\log\xi
    +
    O(1),
    \qquad
    \xi\to0^+ ,
    \label{eq:secV_log_Delta_IR_summary}
\end{equation}
which is integrable at the origin. If either plate coupling vanishes,
$\Delta(\xi)=1$ identically. The explicit coefficient $C_{\rm IR}$ and
the proof of its positivity are given in
App.~\ref{app:infrared_relative_determinant}.

The ultraviolet behaviour has an equally direct physical
interpretation. At large Euclidean frequency, the relative determinant
contains only propagation between the two separated plates. Such propagation is exponentially suppressed, so that the
two-plate factor $\mathcal R(\xi)$ defined in
Eq.~\eqref{eq:secV_R_def} vanishes exponentially
and
\begin{equation*}
    \Delta(\xi)\longrightarrow1,
    \qquad
    \xi\longrightarrow\infty .
\end{equation*}
The interaction energy is therefore ultraviolet finite. The one-plate
self-energy contributions have already been removed by the relative
subtraction in Eq.~\eqref{eq:interaction_energy_subtraction}.

\subsection{Zero-mode projection}
\label{sec:zero_mode_overlap_TGTG}

The exact decomposition
\eqref{eq:zero_mode_resolvent_decomposition} separates the localised
zero-mode fluctuations from the massive fluctuations. To isolate the
contribution carried by the zero mode alone, we now consider the
auxiliary projection
\begin{equation}
    G_{ij}(\xi)
    \longrightarrow
    \frac{\psi_i\psi_j}{\xi^2},
    \qquad
    \psi_i=\psi_0(a_i).
    \label{eq:zero_mode_projected_green}
\end{equation}
At this stage the massive fluctuations are completely omitted. The
resulting construction should therefore be regarded as a pure
zero-mode projection, not yet as the large-separation limit of the full
theory.

For later comparison it is convenient to denote the bare coupling of
plate $i$ to the zero-mode fluctuations by
\begin{equation}
    \alpha_i^{(0)}
    \equiv
    \lambda_i\psi_i^2 .
    \label{eq:alpha_i_delta_plate}
\end{equation}
This quantity contains only the plate strength and the local overlap
with the zero-mode profile; no response of the massive fluctuations is
included.

Under the projection \eqref{eq:zero_mode_projected_green}, the
single-plate determinants become
\begin{align}
    D_{1,0}(\xi)
    &=
    1+\frac{\alpha_1^{(0)}}{\xi^2},
    &
    D_{2,0}(\xi)
    &=
    1+\frac{\alpha_2^{(0)}}{\xi^2},
    \label{eq:one_plate_zero_mode_determinants}
\end{align}
whereas the two-plate determinant is
\begin{equation}
    D_{12,0}(\xi)
    =
    1+
    \frac{
        \alpha_1^{(0)}+\alpha_2^{(0)}
    }{\xi^2}.
    \label{eq:two_plate_zero_mode_determinant}
\end{equation}
The two plates couple here to the same zero-mode fluctuation, so their
bare zero-mode couplings enter the joint determinant additively.

The corresponding relative determinant is therefore
\begin{equation}
    \Delta_{\rm zm}(\xi)
    =
    \frac{
        D_{12,0}(\xi)
    }{
        D_{1,0}(\xi)D_{2,0}(\xi)
    }.
    \label{eq:zero_mode_projected_relative_determinant}
\end{equation}
Substitution into the vacuum-energy integral gives
\begin{equation}
    E_{\rm zm}
    =
    \frac12
    \left[
        \sqrt{
            \alpha_1^{(0)}+\alpha_2^{(0)}
        }
        -
        \sqrt{\alpha_1^{(0)}}
        -
        \sqrt{\alpha_2^{(0)}}
    \right].
    \label{eq:zero_mode_projected_energy}
\end{equation}

Equation~\eqref{eq:zero_mode_projected_energy} isolates the interaction
carried by the pure zero-mode projection. It is not, by itself, the
large-separation asymptotics of the complete interaction, because the
massive fluctuations have been removed altogether. In the full theory
their local response dresses the coupling of each plate to the
zero-mode fluctuations, while their direct propagation between widely
separated plates becomes subleading. In
Sec.~\ref{sec:large_separation_asymptotics} we show explicitly that the
complete relative determinant approaches the same functional form as
Eq.~\eqref{eq:zero_mode_projected_relative_determinant}, with the bare
zero-mode couplings replaced by their dressed infrared values.

\subsection{Scattering interpretation}
\label{sec:scattering_interpretation}

The same relative determinant also admits a direct one-dimensional
scattering interpretation. Let
$\rho^R_{\lambda,a}(\xi)$ and $\rho^L_{\lambda,a}(\xi)$ denote,
respectively, the right- and left-reflection amplitudes of a single
delta plate in the P\"oschl--Teller scattering basis, analytically
continued from real momentum to the imaginary-momentum axis. The
scattering construction and its continuation are derived in
App.~\ref{app:one_dimensional_scattering}.

For $a_1<a_2$, one finds
\begin{equation}
    \rho^R_{\lambda_1,a_1}(\xi)
    \rho^L_{\lambda_2,a_2}(\xi)
    =
    \mathcal R(\xi),
    \label{eq:secVI_reflection_identity}
\end{equation}
where $\mathcal R(\xi)$ is the two-plate scattering factor defined in
Eq.~\eqref{eq:secV_R_def}. Therefore
\begin{equation}
    E_{\rm int}
    =
    \frac{1}{2\pi}
    \int_0^\infty d\xi\,
    \log\left[
        1-
        \rho^R_{\lambda_1,a_1}(\xi)
        \rho^L_{\lambda_2,a_2}(\xi)
    \right].
    \label{eq:secVI_scattering_TGTG}
\end{equation}
Thus the factor that connects the two plates in the
background-dressed $TGTG$ determinant is precisely the product of the
two one-plate reflection amplitudes after analytic continuation.

In a homogeneous background, translational invariance allows the
position dependence of each reflection amplitude to be factored into
free propagation, so the interaction depends only on the separation
$d=a_2-a_1$ and contains the familiar factor $e^{-2\kappa d}$. For the
fixed P\"oschl--Teller background, by contrast, the quadratic
fluctuation problem is not translationally invariant. The reflection
amplitudes therefore depend separately on the plate positions through
the background profile, and the interaction energy depends on their
location relative to the P\"oschl--Teller centre as well as on their
separation.

\section{Limiting regimes and consistency checks}
\label{sec:limiting_regimes}

We now record several limits of the background-dressed $TGTG$
formula derived in Sec.~VI. These checks are useful
both for comparison with the translation-invariant problem
and for the interpretation of the interaction in the
P\"oschl--Teller background. Throughout this section we use
\begin{equation*}
    c=\frac{a_1+a_2}{2},
    \qquad
    d=a_2-a_1>0,
    \qquad
    \kappa=\sqrt{1+\xi^2}.
\end{equation*}

First, the transparent limit is immediate. If either coupling vanishes,
one of the two plates is absent. Since then $\mathcal R(\xi)=0$, the
relative determinant is identically one and
\begin{equation}
	E_{\rm int}=0
	\qquad
	\text{if}
	\qquad
	\lambda_1\lambda_2=0 .
	\label{eq:secVI_transparent_limit}
\end{equation}

The weak-coupling limit is nonuniform because the reference operator
has a zero mode. Setting $\lambda_i=\varepsilon \ell_i$, with fixed
$\ell_i\geq0$, the relative determinant gives, for every fixed
$\xi>0$,
\begin{equation}
    \log\Delta(\xi)
    =
    -\varepsilon^2
    \ell_1\ell_2
    G_{12}(\xi)^2
    +
    O(\varepsilon^3).
    \label{eq:secVI_weak_coupling}
\end{equation}
This expansion cannot be integrated term by term at the infrared
endpoint. Indeed, the zero-mode pole implies
$G_{12}(\xi)\sim\psi_1\psi_2/\xi^2$, so the coefficient of
$\varepsilon^2$ behaves as $\xi^{-4}$ and is not integrable at the
origin. The integrated interaction energy is therefore nonanalytic in
the weak-coupling parameter; the square-root behaviour of
Eq.~\eqref{eq:zero_mode_projected_energy} is the corresponding
zero-mode manifestation.%
\footnote{If the zero mode is lifted, for example by the positive shift
$K_{\rm PT}\to K_{\rm PT}+\mu^2$, the Euclidean Green function is finite
at $\xi=0$ and the conventional termwise weak-coupling expansion is
recovered. The associated stability conditions are discussed in
App.~\ref{app:birman_schwinger}.}

The strong-coupling limit corresponds to imposing Dirichlet conditions
at the two plate positions. Indeed, as $\lambda_i\to+\infty$, the
matching condition \eqref{eq:delta_matching_matrix} forces the normal
modes to vanish at $a_i$. The exact relative determinant then becomes
\[
    \Delta_{\rm D}(\xi)
    =
    1-
    \frac{G_{12}(\xi)^2}
    {G_{11}(\xi)G_{22}(\xi)} .
\]
Using the explicit P\"oschl--Teller Green function, the interaction
energy reduces to
\begin{equation}
\begin{aligned}
E_{\rm D}
&=
\frac{1}{2\pi}
\int_0^\infty d\xi\,
\log\Bigg[
1-e^{-2\kappa d}
\\
&\hspace{-1.5em}
\times
\frac{
\xi^2\cosh(2c)+(2+\xi^2)\cosh d+2\kappa\sinh d
}{
\xi^2\cosh(2c)+(2+\xi^2)\cosh d-2\kappa\sinh d
}
\Bigg].
\end{aligned}
\label{eq:secVI_dirichlet_limit}
\end{equation}
The dependence on the center coordinate $c$ remains explicit through
the factor $\cosh(2c)$. Thus, even in the Dirichlet limit, the
interaction is a genuine function $E_{\rm D}(c,d)$ rather than a
function of the separation $d$ alone.

In a homogeneous massive background,
\[
    G_{11}=G_{22}=\frac{1}{2\kappa},
    \qquad
    G_{12}=\frac{e^{-\kappa d}}{2\kappa},
\]
and the exact relative determinant gives
\begin{equation*}
    E_{\rm int}^{(0)}(d)
    =
    \frac{1}{2\pi}
    \int_0^\infty d\xi\,
    \log\left[
        1-
        \frac{
            \lambda_1\lambda_2 e^{-2\kappa d}
        }{
            (2\kappa+\lambda_1)(2\kappa+\lambda_2)
        }
    \right],
\end{equation*}
the standard massive double-delta result. As required by translational
invariance, it depends only on the separation $d$.

The same result is recovered from the P\"oschl--Teller problem itself by
translating the two plates far from the kink, $|c|\to\infty$ at fixed
$d$. For every fixed $\xi>0$, the exact Green-function entries approach
their homogeneous values. The only nonuniformity is confined to the
shrinking infrared region $\xi=O(e^{-|c|})$ associated with the
zero-mode pole, whose contribution to the interaction energy is
$O(e^{-|c|})$. Therefore
\begin{equation}
    \lim_{|c|\to\infty}E_{\rm int}(c,d)
    =
    E_{\rm int}^{(0)}(d).
    \label{eq:homogeneous_limit}
\end{equation}

\subsection{Dressed zero-mode asymptotics}
\label{sec:large_separation_asymptotics}

We now determine the large-separation behaviour of the full relative
determinant without discarding the massive fluctuations. We consider
$\lambda_1,\lambda_2>0$ and take $d\to\infty$ with $c$, $\lambda_1$
and $\lambda_2$ fixed.

The exact Green-function decomposition introduced in Sec.~VI gives, at
the position of an isolated plate,
\begin{equation}
    G_{ii}(\xi)
    =
    \frac{\psi_i^2}{\xi^2}
    +
    g^\perp_{ii}(\xi).
\end{equation}
Inserting this expression into the exact one-plate determinant
$D_i(\xi)=1+\lambda_iG_{ii}(\xi)$ yields
\begin{equation}
    D_i(\xi)
    =
    1+\lambda_i g^\perp_{ii}(\xi)
    +
    \frac{\alpha_i^{(0)}}{\xi^2},
\end{equation}
where $\alpha_i^{(0)}=\lambda_i\psi_i^2$ is the bare coupling of plate
$i$ to the zero-mode fluctuations introduced in
Sec.~\ref{sec:zero_mode_overlap_TGTG}. Factoring out the local
massive-fluctuation contribution gives the exact identity
\begin{equation}
    D_i(\xi)
    =
    \left[
        1+\lambda_i g^\perp_{ii}(\xi)
    \right]
    \left[
        1+\frac{\alpha_i(\xi)}{\xi^2}
    \right],
    \label{eq:one_plate_zero_massive_factorization}
\end{equation}
where
\begin{equation}
    \alpha_i(\xi)
    =
    \frac{\alpha_i^{(0)}}
    {1+\lambda_i g^\perp_{ii}(\xi)}
    =
    \frac{\lambda_i\psi_i^2}
    {1+\lambda_i g^\perp_{ii}(\xi)} .
    \label{eq:large_sep_dressed_coupling}
\end{equation}
Thus $\alpha_i(\xi)$ follows directly from the exact one-plate
determinant. It is the coupling of plate $i$ to the zero-mode
fluctuations dressed by the local response of the massive
fluctuations.

Since $g^\perp_{ii}(\xi)$ is regular at $\xi=0$, the infrared value of
the dressed coupling is simply $\alpha_i(0)$. For the
P\"oschl--Teller background,
\begin{equation*}
    g^\perp_{ii}(0)
    =
    \frac12-\frac14\operatorname{sech}^2a_i ,
\end{equation*}
so that
\begin{equation}
    \alpha_i(0)
    =
    \frac{
        (\lambda_i/2)\operatorname{sech}^2a_i
    }{
        1+\lambda_i
        \left(
            \frac12-\frac14\operatorname{sech}^2a_i
        \right)
    } .
    \label{eq:large_sep_dressed_alpha_PT}
\end{equation}
At fixed $c$ and large $d$,
\begin{equation}
\begin{aligned}
    \alpha_i(0)
    &=
    \beta_i e^{-d}+O(e^{-2d}),
    \\
    \beta_1
    &=
    \frac{4\lambda_1e^{2c}}{2+\lambda_1},
    \qquad
    \beta_2
    =
    \frac{4\lambda_2e^{-2c}}{2+\lambda_2}.
\end{aligned}
\label{eq:large_sep_beta}
\end{equation}

The second role of the massive fluctuations is their direct propagation
between the plates. After subtraction of the zero-mode contribution,
the exact P\"oschl--Teller Green function gives
$g^\perp_{12}(\xi)=O(de^{-d})$, uniformly on the infrared scale
$\xi^2=O(e^{-d})$. Direct propagation of massive fluctuations is
therefore subleading at large separation, whereas their local response
remains at leading order through the dressed couplings $\alpha_i(0)$.

Since $\alpha_i(0)=O(e^{-d})$, the interaction is controlled by the
infrared scale $\xi=O(e^{-d/2})$. On this scale, regularity of
$g^\perp_{ii}(\xi)$ at the origin gives
\[
    \alpha_i(\xi)
    =
    \alpha_i(0)\,[1+O(\xi^2)] .
\]
The full relative determinant therefore reduces, to leading order, to
\begin{equation}
    \Delta_{\rm IR}(\xi)
    =
    \frac{
        \xi^2
        \left[
            \xi^2+\alpha_1(0)+\alpha_2(0)
        \right]
    }{
        \left[
            \xi^2+\alpha_1(0)
        \right]
        \left[
            \xi^2+\alpha_2(0)
        \right]
    } .
    \label{eq:large_sep_IR_determinant}
\end{equation}
This has the same relative-determinant structure as the pure zero-mode
projection of Sec.~\ref{sec:zero_mode_overlap_TGTG}, but the bare
couplings $\alpha_i^{(0)}$ are replaced by their dressed infrared
values $\alpha_i(0)$.

More precisely, setting $\xi=e^{-d/2}y$ gives the scaled limit
\begin{equation}
\begin{aligned}
    \Delta(e^{-d/2}y)
    &\longrightarrow
    \Delta_\infty(y),
    \\
    \Delta_\infty(y)
    &=
    \frac{
        y^2(y^2+\beta_1+\beta_2)
    }{
        (y^2+\beta_1)(y^2+\beta_2)
    },
    \qquad d\to\infty .
\end{aligned}
\label{eq:large_sep_scaled_limit}
\end{equation}
The passage from this pointwise limit to the energy integral is
controlled directly by the exact factor $\mathcal R$ entering
$\Delta=1-\mathcal R$. Writing
$\mathcal R_d(y)=\mathcal R(e^{-d/2}y)$, the following uniform bound
holds for $d\geq1$:\footnote{The bound follows directly from the exact
expression for $\mathcal R$ after the rescaling
$\xi=e^{-d/2}y$. Writing
$t=\sqrt{1+e^{-d}y^2}-1\geq0$, the relevant $d$-dependent factor obeys
$(1+t)^2e^{-2td}\leq1$ for $d\geq1$.}
\begin{equation}
    0\leq\mathcal R_d(y)
    \leq
    \frac{4\lambda_1\lambda_2}{y^4}.
    \label{eq:large_sep_R_bound}
\end{equation}
At the origin the zero-mode cancellation gives a quadratic zero of
$\Delta$, so that the logarithm is bounded by an integrable
$C+2|\log y|$ behaviour. Dominated convergence therefore applies to
the rescaled energy integral.

Using Eq.~\eqref{eq:interaction_energy_explicit_relative_det}, we obtain
\begin{equation}
    E_{\rm int}(c,d)
    =
    \frac{e^{-d/2}}{2}
    \left[
        \sqrt{\beta_1+\beta_2}
        -
        \sqrt{\beta_1}
        -
        \sqrt{\beta_2}
    \right]
    +
    o(e^{-d/2}).
    \label{eq:large_sep_energy_1p1}
\end{equation}
Thus the large-distance interaction is governed by quantum fluctuations
associated with the localised zero mode. Virtual massive fluctuations remain essential because their local
response dresses the bare zero-mode couplings into $\alpha_i(0)$, while their direct propagation
between the plates is subleading.

The corresponding large-distance force follows from differentiating
the exact rescaled determinant before taking the limit.%
\footnote{After the rescaling $\xi=e^{-d/2}y$, differentiation of the exact
expression for $\mathcal R_d(y)$ at fixed $c$ and $y$ gives a
contribution that vanishes as $d\to\infty$. The zero-mode cancellation keeps the
derivative of the logarithm integrable at $y=0$, while the
large-$y$ behaviour is $O(y^{-4})$ uniformly for sufficiently large
$d$. Dominated convergence therefore permits the derivative to be
taken through the rescaled integral.}
Defining $F_d=-\partial_dE_{\rm int}$, one finds
\begin{equation}
    F_d(c,d)
    =
    \frac{e^{-d/2}}{4}
    \left[
        \sqrt{\beta_1+\beta_2}
        -
        \sqrt{\beta_1}
        -
        \sqrt{\beta_2}
    \right]
    +
    o(e^{-d/2}).
    \label{eq:large_sep_force_1p1}
\end{equation}

The central point is therefore that the full determinant does not
approach the bare zero-mode projection. It approaches the same zero-mode relative-determinant structure with
the locally dressed infrared couplings $\alpha_i(0)$. The massive fluctuations survive at
leading order through this local dressing, whereas their direct
propagation between widely separated plates is exponentially
subleading.

\section{Position-dependent interaction and forces}
\label{sec:position_dependent_interaction}

The derivation of Sec.~\ref{sec:relative_determinant_TGTG} was carried
out in one spatial dimension because this is where the nontrivial part
of the problem lies. The Pöschl--Teller background, the point-supported
interactions, the relative determinant and the stability analysis are all
encoded in the transverse one-dimensional operator. The extension to
planar plates in $(3+1)$ dimensions is then obtained by adding two free
directions parallel to the plates. After Wick rotation, a momentum $\mathbf{k}_\parallel$ parallel to
the plates enters the one-dimensional problem through
\begin{equation*}
	p=\sqrt{\xi^2+\mathbf{k}_\parallel^2}.
\end{equation*}
Hence the relative determinant has the same functional form as in one
spatial dimension, with the Euclidean frequency replaced by $p$,
$\Delta(\xi)\to\Delta(p)$. For the reference Pöschl--Teller
background, the zero mode remains localized in the transverse direction;
the corresponding massless fluctuations propagate only along the
directions parallel to the plates.

Integrating over $\xi$ and $\mathbf{k}_\parallel$, and writing $S$ for
the plate area, gives
\begin{equation}
	\frac{E_{\rm int}}{S}
	=
	\frac{1}{4\pi^2}
	\int_0^\infty dp\,p^2\log\Delta(p).
	\label{eq:secVIII_energy_density_3p1}
\end{equation}
Using the center coordinate $c$ and plate separation $d$ introduced in
Sec.~\ref{sec:limiting_regimes}, the derivatives of the relative
interaction energy define
\begin{equation}
	\frac{F_d}{S}
	=
	-\frac{\partial}{\partial d}
	\frac{E_{\rm int}}{S},
	\qquad
	\frac{F_c}{S}
	=
	-\frac{\partial}{\partial c}
	\frac{E_{\rm int}}{S}.
	\label{eq:secVIII_force_densities}
\end{equation}
Here $F_d$ is the interaction force conjugate to the plate separation,
whereas $F_c$ is the interaction contribution associated with a rigid
displacement of the plate pair relative to the fixed background.
For $\lambda_1,\lambda_2>0$, differentiation with respect to $c$ and
$d$ may be interchanged with the integration in
Eq.~\eqref{eq:secVIII_energy_density_3p1}. Near $p=0$,
Eq.~\eqref{eq:secV_log_Delta_IR_summary} gives
$\log\Delta(p)=2\log p+O(1)$. The singular term is independent of $c$
and $d$, while differentiation of the remaining part is regular at the
origin. At large $p$, differentiation with respect to either parameter
preserves the exponential suppression generated by the off-diagonal
Green function. The differentiated integrands are therefore integrable
at both endpoints. If either coupling vanishes, $\Delta(p)=1$
identically and both interaction-force densities vanish.

For identical plates, the reflection symmetry of the Pöschl--Teller
background gives an exact constraint on the center-coordinate
dependence. Under
$(a_1,a_2)\mapsto(-a_2,-a_1)$ one has $c\mapsto-c$ while $d$
remains unchanged. Since the background is even,
\begin{equation}
	E_{\rm int}(c,d;\lambda,\lambda)
	=
	E_{\rm int}(-c,d;\lambda,\lambda).
	\label{eq:secVIII_reflection_symmetry}
\end{equation}
Therefore the center-force contribution is odd in $c$,
\begin{equation}
	F_c(c,d;\lambda,\lambda)
	=
	-F_c(-c,d;\lambda,\lambda),
	\label{eq:secVIII_center_force_odd}
\end{equation}
and in particular
\begin{equation}
	F_c(0,d;\lambda,\lambda)=0.
	\label{eq:secVIII_center_force_zero}
\end{equation}
This exact symmetry will provide a useful check on the numerical
results below.

For $\lambda_1,\lambda_2>0$ with unequal couplings, the center-force
contribution at $c=0$ is generically nonzero. Its sign follows directly
from the exact determinant. At $c=0$, reflection symmetry gives
\[
	G_{11}=G_{22},
	\qquad
	\partial_cG_{12}=0,
	\qquad
	\partial_cG_{11}=-\partial_cG_{22}>0.
\]
Equation~\eqref{eq:secV_R_def} then yields
\[
	\left.
	\partial_c\log\mathcal R
	\right|_{c=0}
	=
	\frac{
		(\lambda_2-\lambda_1)\,\partial_cG_{11}
	}{
		(1+\lambda_1G_{11})(1+\lambda_2G_{11})
	}.
\]
Since $\Delta=1-\mathcal R$ with $0<\mathcal R<1$, it follows that
\[
	\operatorname{sgn}F_c(0,d)
	=
	\operatorname{sgn}(\lambda_2-\lambda_1),
	\qquad
	\lambda_1,\lambda_2>0.
\]

The large-separation analysis of
Sec.~\ref{sec:limiting_regimes} immediately gives the corresponding
planar asymptotics. With
$p=e^{-d/2}y$, the same scaled-determinant limit
\eqref{eq:large_sep_scaled_limit} applies, while
$p^2dp=e^{-3d/2}y^2dy$. The bound
\eqref{eq:large_sep_R_bound} remains integrable after multiplication
by $y^2$, so no additional asymptotic argument is required. Hence
\begin{equation}
	\frac{E_{\rm int}}{S}
	=
	\frac{e^{-3d/2}}{12\pi}
	\left[
		\beta_1^{3/2}
		+
		\beta_2^{3/2}
		-
		(\beta_1+\beta_2)^{3/2}
	\right]
	+
	o(e^{-3d/2}).
	\label{eq:large_sep_energy_3p1}
\end{equation}
Here the $\beta_i$ are those defined in
Eq.~\eqref{eq:large_sep_beta}. The change from the
$e^{-d/2}$ law in $(1+1)$ dimensions to $e^{-3d/2}$ in the planar
problem is therefore entirely due to the Euclidean phase-space measure,
not to a different relative determinant.

The differentiation with respect to $d$ can be justified as in
Sec.~\ref{sec:limiting_regimes}. After the rescaling
$p=e^{-d/2}y$, the infrared behaviour remains integrable, while at
large $y$ the derivative of the logarithm is $O(y^{-4})$; the planar
measure therefore gives an $O(y^{-2})$ integrand.

The derivative with respect to $c$ can be justified directly as well. At fixed
$d$, it acts only on the smooth $c$ dependence of the exact
Green-function entries. The quadratic infrared behaviour of
$\Delta$ implies
$\partial_c\log\Delta=O(1)$ as $y\to0$, whereas for large $y$ one
again has $\partial_c\log\Delta=O(y^{-4})$. Thus the corresponding
planar integrand is integrable at both endpoints. The large-separation
interaction-force densities are therefore
\begin{equation}
	\frac{F_d}{S}
	=
	\frac{e^{-3d/2}}{8\pi}
	\left[
		\beta_1^{3/2}
		+
		\beta_2^{3/2}
		-
		(\beta_1+\beta_2)^{3/2}
	\right]
	+
	o(e^{-3d/2}),
	\label{eq:large_sep_Fd_3p1}
\end{equation}
and
\begin{align}
	\frac{F_c}{S}
	={}&
	-\frac{e^{-3d/2}}{4\pi}
	\Big[
		\beta_1^{3/2}
		-
		\beta_2^{3/2}
		\nonumber\\
	&\hspace{1.7cm}
	-
		(\beta_1-\beta_2)
		\sqrt{\beta_1+\beta_2}
	\Big]
	+
	o(e^{-3d/2}).
	\label{eq:large_sep_Fc_3p1}
\end{align}

The large-separation expression for $F_c$ also determines how the
asymmetry between the two plate couplings shifts the stationary point
of the relative interaction energy. Since unequal couplings generally
give a nonzero center-force contribution at $c=0$, we set the leading
term of Eq.~\eqref{eq:large_sep_Fc_3p1} to zero. For
$\lambda_1,\lambda_2>0$, this is equivalent to $\beta_1=\beta_2$ and
gives
\[
c_*^{(\infty)}
=
\frac14
\log\left[
\frac{\lambda_2(2+\lambda_1)}
{\lambda_1(2+\lambda_2)}
\right],
\]
with
\[
\operatorname{sgn}c_*^{(\infty)}
=
\operatorname{sgn}(\lambda_2-\lambda_1).
\]
Thus, at leading order for $d\to\infty$, unequal plate couplings shift
the stationary point of the relative interaction energy away from the
center of the Pöschl--Teller background.

The numerical evaluation below uses the exact planar expression
\eqref{eq:secVIII_energy_density_3p1} with the full relative determinant
\eqref{eq:secV_Delta_def}. It is therefore neither a zero-mode
projection nor a large-separation approximation. The numerical results
include the zero-mode fluctuations, the massive fluctuations, and all
multiple-scattering effects. Equations
\eqref{eq:large_sep_energy_3p1}--\eqref{eq:large_sep_Fc_3p1} provide
instead the controlled large-separation asymptotics of the same exact
interaction.

\begin{figure*}[t]
	\centering
	\includegraphics[width=0.98\textwidth]{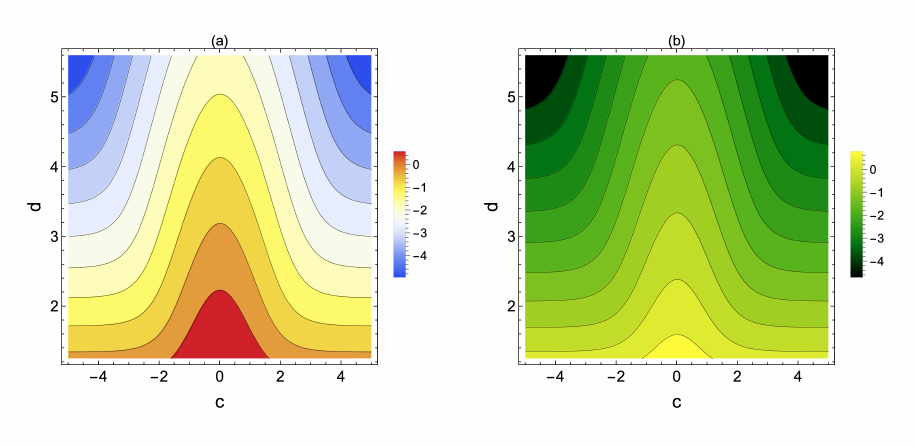}
	\caption{
    Interaction energy density and interplate force density for two
    identical repulsive plates, $\lambda_1=\lambda_2=5$, in the planar
    $(3+1)$-dimensional extension. Panel (a) shows
    $10^3 E_{\rm int}/S$ and panel (b) $10^3 F_d/S$;
    geometrically spaced contour levels are used in both panels.
    The interaction remains attractive in the region shown, but
    both quantities depend on the center coordinate $c$, not only on
    the separation $d$.
}
	\label{fig:secVIII_energy_Fd}
\end{figure*}
\begin{figure}[t]
	\centering
	\includegraphics[width=0.88\columnwidth]{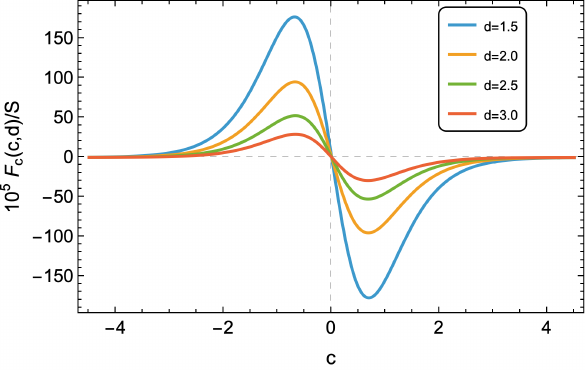}
	\caption{
		Interaction center-force density $10^5F_c/S$ as a function
		of $c$ for two identical plates, $\lambda_1=\lambda_2=5$, and
		several values of $d$. Reflection symmetry forces $F_c=0$ at
		$c=0$. Away from the center, the sign of $F_c$ shows that the
		relative interaction energy generates a center-coordinate
		contribution directed toward the Pöschl--Teller well.
	}
	\label{fig:secVIII_Fc_symmetric}
\end{figure}
\begin{figure}[t]
	\centering
	\includegraphics[width=0.88\columnwidth]{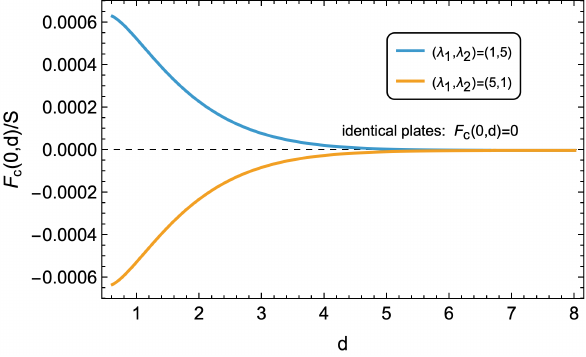}
	\caption{
		Interaction center-force density $F_c/S$ at $c=0$ for
		asymmetric plates. The two curves correspond to
		$(\lambda_1,\lambda_2)=(1,5)$ and $(5,1)$. Unlike the
		identical-plate case, the center-force contribution is generically
		nonzero: geometrical symmetry does not imply stationarity of the
		relative interaction energy when the transparencies of the plates are
		different.
	}
	\label{fig:secVIII_Fc_asymmetric}
\end{figure}

Figure~\ref{fig:secVIII_energy_Fd} shows the interaction energy density
and the interplate force density for two identical repulsive plates. The interaction energy is not a function of
$d$ alone: its magnitude depends strongly on the location of the pair
relative to the P\"oschl--Teller center. This is the most direct
numerical signature of the loss of translational invariance of the
fixed-background quadratic fluctuation theory, encoded in the relative
determinant through the entries $G_{ij}$ defined in
Eq.~\eqref{eq:secIV_Gij_def}. Since the calculation uses the full
relative determinant, it includes both the zero-mode and massive
fluctuations, together with all multiple-scattering effects. The force
$F_d$ remains negative throughout the parameter region shown, so the
interaction between the two repulsive plates remains attractive there.
Within this region, the background strongly modulates the magnitude of
the force without changing its sign.

The center-coordinate derivative in
Eq.~\eqref{eq:secVIII_force_densities}, $F_c/S$, has no analogue in
the free translation-invariant double-delta problem. The oddness in
Eq.~\eqref{eq:secVIII_center_force_odd} is visible in
Fig.~\ref{fig:secVIII_Fc_symmetric}. Away from $c=0$, however, this
interaction contribution is nonzero. Thus the relative interaction energy
depends on the rigid position of the plate pair with respect to the
fixed Pöschl--Teller background.

For the identical plates shown in
Figs.~\ref{fig:secVIII_energy_Fd} and
\ref{fig:secVIII_Fc_symmetric}, the interaction energy becomes more
negative as $|c|$ decreases at fixed $d$. Accordingly, the interaction
contribution $F_c=-\partial_cE_{\rm int}$ is directed toward $c=0$.
This behaviour follows from the separate dependence of the exact
Green-function entries $G_{ij}$ on the two plate positions.

For plates with unequal couplings, the center-force contribution at
$c=0$ need not vanish, so a geometrically symmetric placement is not,
in general, a stationary point of the relative interaction energy.
Figure~\ref{fig:secVIII_Fc_asymmetric} illustrates this behaviour.
The sign of $F_c(0,d)$ reverses when the two couplings are interchanged,
in agreement with the exact result derived above.

The nonzero center-coordinate contribution is a direct consequence of
the fixed Pöschl--Teller background. It disappears in the homogeneous
limit, where translational invariance reduces the relative interaction
energy to a function of the plate separation alone.

The quantities shown in
Figs.~\ref{fig:secVIII_Fc_symmetric} and
\ref{fig:secVIII_Fc_asymmetric} are derivatives of the relative
interaction energy only. The position-dependent one-plate self-energies
subtracted in its definition would contribute separately to the total
mechanical force and are not included here.

The numerical results should be read together with the controlled
large-separation limit derived above. The full relative determinant does
not approach the bare zero-mode projection. Instead, it approaches the
same zero-mode relative-determinant structure with the dressed infrared
couplings $\alpha_i(0)$. The massive fluctuations remain relevant at leading order
through their local response entering $\alpha_i(0)$, whereas their direct
propagation between widely separated plates is subleading.

Finally, it is useful to restore the physical scales. Since the
dimensionless transverse coordinate is defined by $x=m\widetilde x$,
the mass parameter $m$ sets the inverse length scale of the
Pöschl--Teller profile. Thus
\[
	\widetilde c=\frac{c}{m},
	\qquad
	\widetilde d=\frac{d}{m},
\]
and values $c,d=O(1)$ correspond to distances of the order of the
characteristic wall width $m^{-1}$. The dimensionful interaction energy
is $mE_{\rm int}$ in $(1+1)$ dimensions, whereas in the planar
$(3+1)$-dimensional extension the interaction energy per unit area is
$m^3E_{\rm int}/S$. The position dependence displayed in
Figs.~\ref{fig:secVIII_energy_Fd}--\ref{fig:secVIII_Fc_asymmetric}
therefore probes the regime in which the plates lie at distances
comparable to the characteristic width of the fixed Pöschl--Teller
background.

\section{Conclusions}
\label{sec:conclusions}

In this paper we have derived an exact background-dressed $TGTG$ formula
for the one-loop interaction energy of two semitransparent delta plates
embedded in the one-soliton P\"oschl--Teller background associated with
the sine-Gordon kink. The construction is based on a relative determinant
in which the two isolated one-plate contributions are subtracted in the
same P\"oschl--Teller background. The resulting interaction energy is
therefore separated from the corresponding one-body self-energies from
the outset.

The relative determinant follows from the finite-dimensional
boundary-space Birman--Schwinger reduction. Both the one-plate transition
operators and the propagation between the plates are defined with respect
to the P\"oschl--Teller resolvent. The result is therefore not obtained by
formally replacing the free Green function by the P\"oschl--Teller Green
function in the translation-invariant double-delta formula. This
same-background construction is the precise sense in which the $TGTG$
structure is background dressed. The exactly solvable P\"oschl--Teller
system thus provides a controlled prototype of such a relative-determinant
construction in an inhomogeneous background.

The resulting interaction energy depends on the two plate positions
separately. Equivalently, in terms of the center coordinate $c$ and the
separation $d$, it is a genuine function $E_{\rm int}(c,d)$ rather than a
function of $d$ alone. This position dependence is a direct consequence
of the fixed-background quadratic fluctuation problem: once the kink is
held fixed, the P\"oschl--Teller profile selects a distinguished spatial
region and the fluctuation theory is no longer translationally invariant.
This does not imply a loss of translational invariance of the underlying
sine-Gordon theory, whose translational zero mode is retained explicitly
in the fluctuation determinant.

We have also analyzed the stability of the fluctuation problem. For $\lambda_1,\lambda_2\geq0$, the repulsive delta interactions cannot
drive any fluctuation mode to negative $\omega^2$, so the fluctuation QFT
remains in the stable sector. By
contrast, an isolated attractive plate destabilizes the zero mode in the
unshifted P\"oschl--Teller background. Since the relative determinant
subtracts the two isolated one-plate configurations, the unshifted
construction is therefore restricted to $\lambda_1,\lambda_2\geq0$,
independently of whether a particular combined two-plate configuration
might remain stable. To include attractive plates within the relative construction, the
reference problem must be modified. A positive mass shift lifts the zero
mode and provides such a modification, with the admissible coupling
domain determined by the corresponding one- and two-plate
Birman--Schwinger conditions. Within this domain no negative normal modes,
and hence no imaginary-frequency instabilities, are present.

The zero-mode pole of the P\"oschl--Teller resolvent does not spoil the
interaction formula. Although the Green function is singular at zero
Euclidean frequency, the relative determinant produces only an integrable
logarithmic singularity. The pure zero-mode projection is characterized by
the bare couplings
$\alpha_i^{(0)}=\lambda_i\psi_i^2$, but this projection is not by itself
the large-separation limit of the complete interaction. In the full
theory, the local response of the massive fluctuations dresses these
couplings according to
\[
	\alpha_i(\xi)
	=
	\frac{\alpha_i^{(0)}}{1+\lambda_i g^\perp_{ii}(\xi)} .
\]
At large plate separation, the interaction is controlled by the infrared
scale $\xi=O(e^{-d/2})$, and the full relative determinant approaches the
same zero-mode relative-determinant structure with the dressed infrared
couplings $\alpha_i(0)$. The direct propagation of massive fluctuations
between the plates is then subleading, while their local response remains
present at leading order through $\alpha_i(0)$. In this exactly solvable setting, the complete one-loop calculation
therefore shows explicitly how the infrared zero-mode mechanism discussed
in Ref.~\cite{MunozCastanedaTopologicalCasimir_inprep} emerges from the
full relative determinant. Correspondingly, the large-separation interaction behaves as
$E_{\rm int}=O(e^{-d/2})$ in $(1+1)$ dimensions and
$E_{\rm int}/S=O(e^{-3d/2})$ in the planar $(3+1)$-dimensional extension.

In the planar $(3+1)$-dimensional extension we have analysed the
derivatives of the relative interaction energy with respect to the plate
separation $d$ and the center coordinate $c$. For the repulsive
configurations explored numerically, the interplate force remains
attractive, although its magnitude is strongly modulated by the
P\"oschl--Teller background. The center-coordinate derivative provides
an interaction contribution with no analogue in the homogeneous
translation-invariant problem. For identical plates, reflection symmetry
makes this contribution odd in $c$ and therefore gives
$F_c(0,d)=0$ exactly. For unequal strictly repulsive plates, the exact
relative determinant gives
\[
	\operatorname{sgn}F_c(0,d)
	=
	\operatorname{sgn}(\lambda_2-\lambda_1).
\]
At large separation, the leading asymptotic interaction locates the
stationary point of the relative interaction energy away from the center
of the P\"oschl--Teller background when the plate couplings are unequal.

This center-coordinate contribution should not be confused with the
total mechanical force associated with a rigid displacement of the
two-plate system. The relative determinant isolates the mutual
interaction by subtracting the two isolated one-plate vacuum energies
in the same P\"oschl--Teller background. In this inhomogeneous
background those one-plate self-energies are themselves
position dependent and would therefore contribute separately to the
total force. Their analysis, including the corresponding
renormalization problem, constitutes a distinct one-loop problem and
is the subject of ongoing work. Accordingly, the present relative
interaction calculation does not by itself establish mechanical
trapping of the plates. A dynamical treatment of the translational
collective coordinate of the kink would additionally be required to
describe the reciprocal backreaction on the wall and the complete
force balance.

Several further extensions are natural. One may consider other
reflectionless backgrounds, multi-soliton or periodic backgrounds,
modify the reference problem so as to admit stable attractive plates,
or replace the delta plates by more general point-supported
interactions such as $\delta$--$\delta'$ defects. A positive mass shift
provides one explicit route to the attractive sector, with stability
controlled by the corresponding Birman--Schwinger conditions. More generally, the relative-determinant viewpoint developed here provides
a controlled starting point for analogous $TGTG$ constructions in other
inhomogeneous backgrounds, while keeping the reference problem fixed
and separating mutual interaction energies from position-dependent
one-body contributions.

\appendix


\section{Construction of the Pöschl--Teller Green function}
\label{app:PT_green_function}

In this appendix we derive the Euclidean Green function used in
Sec.~\ref{sec:PT_green_function}. We keep the notation of the main text:
$K_{\rm PT}$ denotes the dimensionless background operator,
$G_\xi^{\rm PT}$ its Euclidean resolvent kernel, and $\xi>0$ the
Euclidean frequency.

The operator is
\begin{equation*}
	K_{\rm PT}
	=
	-\frac{d^2}{dx^2}
	+
	1
	-
	2\,\operatorname{sech}^2 x .
\end{equation*}
For $\xi>0$, the Green function is the integral kernel of
$(K_{\rm PT}+\xi^2)^{-1}$. We set
\begin{equation*}
	\kappa=\sqrt{1+\xi^2},
\end{equation*}
as in Eq.~\eqref{eq:kappa_definition}. The homogeneous equation
associated with the resolvent problem is then
\begin{equation}
	\left[
	-\frac{d^2}{dx^2}
	+
	\kappa^2
	-
	2\,\operatorname{sech}^2 x
	\right]f(x)=0 .
	\label{eq:appA_homogeneous_eq}
\end{equation}

It is useful to obtain its solutions from the elementary factorization
of the one-soliton Pöschl--Teller operator. Define
\begin{equation*}
	A=\frac{d}{dx}+\tanh x,
	\qquad
	A^\dagger=-\frac{d}{dx}+\tanh x .
\end{equation*}
Then
\begin{equation*}
	K_{\rm PT}=A^\dagger A,
	\qquad
	-\frac{d^2}{dx^2}+1=AA^\dagger .
\end{equation*}
The intertwining relation
\begin{equation*}
	K_{\rm PT}A^\dagger
	=
	A^\dagger
	\left(
	-\frac{d^2}{dx^2}+1
	\right)
\end{equation*}
shows that applying $A^\dagger$ to the free solutions produces
solutions of Eq.~\eqref{eq:appA_homogeneous_eq}.

At real momentum one would obtain the usual scattering solutions by
applying the intertwining operator to the plane waves $e^{\pm ikx}$.
However, the vacuum energy formula used in the main text is written
after Wick rotation to imaginary frequency. Thus the relevant objects
in this appendix are not oscillatory scattering waves, but their
Euclidean continuations. Equivalently, the real momentum $k$ has been
continued to $i\kappa$, and the plane waves have become real exponentials.
These Euclidean scattering solutions are selected by their decay at one
of the two spatial infinities.

Since $e^{\pm\kappa x}$ solve
\begin{equation*}
	\left(
	-\frac{d^2}{dx^2}+\kappa^2
	\right)e^{\pm\kappa x}=0,
\end{equation*}
we choose the two Euclidean continuations of the scattering solutions as
\begin{align}
	f_\xi^-(x)
	&=
	e^{\kappa x}\left(\kappa-\tanh x\right),
	\label{eq:appA_fminus}
	\\
	f_\xi^+(x)
	&=
	e^{-\kappa x}\left(\kappa+\tanh x\right).
	\label{eq:appA_fplus}
\end{align}
The normalization is chosen for later convenience. The solution
$f_\xi^-$ decays as $x\to-\infty$, whereas $f_\xi^+$ decays as
$x\to+\infty$. In this Euclidean setting, the terms ``incoming'',
``outgoing'', and ``reflection'' should be understood only after
analytic continuation back to real momentum; here we use only the
decaying solutions needed to construct the resolvent.

With the Wronskian convention
\begin{equation*}
	W[f,g]=f g'-f'g,
\end{equation*}
one finds
\begin{equation}
	W_\xi
	=
	W[f_\xi^-,f_\xi^+]
	=
	-2\kappa(\kappa^2-1).
	\label{eq:appA_W}
\end{equation}
Indeed, the Wronskian is independent of $x$, so it can be evaluated
at $x\to+\infty$. In this limit,
\begin{equation*}
	f_\xi^-(x)\sim(\kappa-1)e^{\kappa x},
	\qquad
	f_\xi^+(x)\sim(\kappa+1)e^{-\kappa x},
\end{equation*}
and therefore
\begin{align*}
	W_\xi
	&=
	(\kappa-1)(\kappa+1)
	W[e^{\kappa x},e^{-\kappa x}]
	\\
	&=
	-2\kappa(\kappa^2-1).
\end{align*}
The vanishing of $W_\xi$ at $\xi=0$ is the analytic signal of the
zero mode of $K_{\rm PT}$.

The Euclidean Green function is built from these two decaying
Euclidean solutions. For $x\neq y$ it solves the homogeneous equation,
while decay at both spatial infinities fixes the form
\begin{equation}
	G_\xi^{\rm PT}(x,y)
	=
	C_\xi\,
	f_\xi^-(x_<)f_\xi^+(x_>),
	\label{eq:appA_green_ansatz}
\end{equation}
where
\begin{equation*}
	x_< = \min\{x,y\},
	\qquad
	x_> = \max\{x,y\}.
\end{equation*}
The constant $C_\xi$ is determined by the normalization convention
used in the main text,
\begin{equation}
	\left(K_{\rm PT}+\xi^2\right)
	G_\xi^{\rm PT}(x,y)
	=
	\delta(x-y).
	\label{eq:appA_green_eq}
\end{equation}
Integrating Eq.~\eqref{eq:appA_green_eq} across $x=y$ gives
\begin{equation}
	\partial_xG_\xi^{\rm PT}(y^-,y)
	-
	\partial_xG_\xi^{\rm PT}(y^+,y)
	=
	1.
	\label{eq:appA_jump_condition}
\end{equation}
Substitution of Eq.~\eqref{eq:appA_green_ansatz} into
Eq.~\eqref{eq:appA_jump_condition} yields
\begin{equation*}
	C_\xi=-\frac{1}{W_\xi}.
\end{equation*}
Hence
\begin{equation*}
	G_\xi^{\rm PT}(x,y)
	=
	-\frac{
		f_\xi^-(x_<)f_\xi^+(x_>)
	}{
		W_\xi
	}.
\end{equation*}

Using Eqs.~\eqref{eq:appA_fminus}, \eqref{eq:appA_fplus}, and
\eqref{eq:appA_W}, we obtain
\begin{equation}
	G_\xi^{\rm PT}(x,y)
	=
	\frac{
		e^{-\kappa |x-y|}
		(\kappa-\tanh x_<)(\kappa+\tanh x_>)
	}{
		2\kappa(\kappa^2-1)
	}.
	\label{eq:appA_green_explicit}
\end{equation}
This is Eq.~\eqref{eq:PT_green_explicit_xi} of the main text. Its
diagonal value is
\begin{equation*}
	G_\xi^{\rm PT}(x,x)
	=
	\frac{
		\kappa^2-\tanh^2x
	}{
		2\kappa(\kappa^2-1)
	}
	=
	\frac{
		\xi^2+\operatorname{sech}^2x
	}{
		2\kappa\xi^2
	}.
\end{equation*}

Finally, we extract the zero-mode pole explicitly. Suppose first that
$x\leq y$. Then $x_<=x$ and $x_>=y$, and
\begin{equation*}
	1-\tanh x
	=
	\frac{e^{-x}}{\cosh x},
	\qquad
	1+\tanh y
	=
	\frac{e^{y}}{\cosh y}.
\end{equation*}
Therefore
\begin{equation*}
	e^{-(y-x)}
	(1-\tanh x)(1+\tanh y)
	=
	\operatorname{sech}x\,\operatorname{sech}y .
\end{equation*}
The same identity follows by symmetry when $y\leq x$. Since
$\kappa=1+O(\xi^2)$ and $\kappa^2-1=\xi^2$, Eq.~\eqref{eq:appA_green_explicit}
gives
\begin{equation*}
	G_\xi^{\rm PT}(x,y)
	=
	\frac{
		\operatorname{sech}x\,\operatorname{sech}y
	}{
		2\xi^2
	}
	+
	O(1),
	\qquad
	\xi\to0^+ .
\end{equation*}
Equivalently, using the normalized zero mode
\begin{equation*}
	\psi_0(x)=\frac{1}{\sqrt2}\operatorname{sech}x,
\end{equation*}
one obtains
\begin{equation*}
	G_\xi^{\rm PT}(x,y)
	=
	\frac{\psi_0(x)\psi_0(y)}{\xi^2}
	+
	O(1),
	\qquad
	\xi\to0^+ .
\end{equation*}
This is the infrared singularity used in
Secs.~\ref{sec:PT_green_function} and
\ref{sec:relative_determinant_TGTG}.

\section{Rank-two determinant identities}
\label{app:rank_two_determinants}

In this appendix we justify the finite-dimensional determinant identities
used in Secs.~\ref{sec:finite_rank_plates}
and~\ref{sec:relative_determinant_TGTG}. The discussion is at
fixed Euclidean frequency $\xi>0$. Let
\begin{equation*}
	\mathcal H=L^2(\mathbb R),
	\qquad
	R_\xi
	=
	G_\xi^{\rm PT}
	=
	\left(K_{\rm PT}+\xi^2\right)^{-1}.
	\label{eq:appB_G_def}
\end{equation*}
The form domain of $K_{\rm PT}$ is $H^1(\mathbb R)$. We introduce the
bounded trace map
\begin{equation*}
	\Gamma:H^1(\mathbb R)\longrightarrow\mathbb C^2,
	\qquad
	\Gamma\psi
	=
	\begin{pmatrix}
		\psi(a_1)\\
		\psi(a_2)
	\end{pmatrix},
\end{equation*}
together with the coupling matrix
\begin{equation*}
	\Lambda
	=
	\begin{pmatrix}
		\lambda_1&0\\
		0&\lambda_2
	\end{pmatrix}.
\end{equation*}
The two-plate perturbation is represented at the quadratic-form level by
\begin{equation*}
	q_\delta[\psi]
	=
	\langle\Gamma\psi,\Lambda\Gamma\psi\rangle_{\mathbb C^2}.
\end{equation*}
The customary notation
\begin{equation*}
	V
	=
	\lambda_1|a_1\rangle\langle a_1|
	+
	\lambda_2|a_2\rangle\langle a_2|
	\label{eq:appB_V_def}
\end{equation*}
is only a distributional shorthand for this form perturbation.

Since $R_\xi^{1/2}$ maps $\mathcal H$ boundedly into
$H^1(\mathbb R)$, the operator
\begin{equation*}
	A_\xi
	=
	\Gamma R_\xi^{1/2}
	:
	\mathcal H\longrightarrow\mathbb C^2
\end{equation*}
is bounded. The associated Birman--Schwinger operator
\begin{equation*}
	B_\xi
	=
	A_\xi^\dagger\Lambda A_\xi
\end{equation*}
has rank at most two and is therefore trace class on $\mathcal H$.
The Fredholm determinant denoted formally by
$\det(1+G_\xi^{\rm PT}V)$ is understood as
\begin{equation*}
	\det_{\mathcal H}
	\left(
		\mathbb I+B_\xi
	\right).
\end{equation*}
Moreover,
\begin{equation*}
	A_\xi A_\xi^\dagger
	=
	\Gamma R_\xi\Gamma^\dagger
	=
	\mathbb G(\xi),
\end{equation*}
where
\begin{equation*}
	\mathbb G(\xi)
	=
	\begin{pmatrix}
		G_{11}(\xi)&G_{12}(\xi)\\
		G_{21}(\xi)&G_{22}(\xi)
	\end{pmatrix}.
\end{equation*}

Sylvester's determinant identity gives
\begin{equation}
	\det_{\mathcal H}
	\left(
		\mathbb I+B_\xi
	\right)
	=
	\det_{\mathbb C^2}
	\left(
		\mathbb I_2+\Lambda A_\xi A_\xi^\dagger
	\right)
	=
	\det_{\mathbb C^2}
	\left(
		\mathbb I_2+\Lambda\mathbb G(\xi)
	\right).
	\label{eq:appB_finite_rank_det_identity}
\end{equation}
Explicitly,
\begin{equation}
	\begin{split}
	D_{12}(\xi)
	&\equiv
	\det_{\mathcal H}
	\left(
		\mathbb I+B_\xi
	\right)
	\\
	&=
	\det_{\mathbb C^2}
	\begin{pmatrix}
		1+\lambda_1G_{11}
		&
		\lambda_1G_{12}
		\\
		\lambda_2G_{21}
		&
		1+\lambda_2G_{22}
	\end{pmatrix}
	\\
	&=
	(1+\lambda_1G_{11})(1+\lambda_2G_{22})
	-
	\lambda_1\lambda_2G_{12}G_{21}.
	\end{split}
	\label{eq:appB_two_plate_det}
\end{equation}
For a single plate, define the trace map
\begin{equation*}
	\Gamma_i:H^1(\mathbb R)\longrightarrow\mathbb C,
	\qquad
	\Gamma_i\psi=\psi(a_i),
\end{equation*}
and set
\begin{equation*}
	A_{i,\xi}
	=
	\Gamma_iR_\xi^{1/2},
	\qquad
	B_{i,\xi}
	=
	\lambda_iA_{i,\xi}^\dagger A_{i,\xi},
	\qquad
	i=1,2.
\end{equation*}
The operator $B_{i,\xi}$ has rank at most one. Hence
\begin{equation}
	D_i(\xi)
	\equiv
	\det_{\mathcal H}
	\left(
		\mathbb I+B_{i,\xi}
	\right)
	=
	1+\lambda_iG_{ii}(\xi),
	\qquad
	i=1,2.
	\label{eq:appB_single_plate_det}
\end{equation}
Therefore the relative determinant that removes the two isolated
one-plate contributions in the same Pöschl--Teller background is
\begin{equation*}
	\Delta(\xi)
	=
	\frac{
		D_{12}(\xi)
	}{
		D_1(\xi)D_2(\xi)
	}.
	\label{eq:appB_Delta_def}
\end{equation*}
Using Eqs.~\eqref{eq:appB_two_plate_det} and
\eqref{eq:appB_single_plate_det}, one obtains
\begin{equation}
	\Delta(\xi)
	=
	1-
	\frac{
		\lambda_1\lambda_2G_{12}G_{21}
	}{
		(1+\lambda_1G_{11})(1+\lambda_2G_{22})
	}.
	\label{eq:appB_relative_det}
\end{equation}
From the explicit Pöschl--Teller Green function in
Eq.~\eqref{eq:PT_green_explicit_xi},
\begin{equation*}
	G_{12}=G_{21}.
	\label{eq:appB_G_symmetry}
\end{equation*}
Thus Eq.~\eqref{eq:appB_relative_det} reduces to
\begin{equation*}
	\Delta(\xi)
	=
	1-
	\frac{
		\lambda_1\lambda_2G_{12}^2
	}{
		(1+\lambda_1G_{11})(1+\lambda_2G_{22})
	},
	\label{eq:appB_relative_det_symmetric}
\end{equation*}
which is the determinant entering the background-dressed $TGTG$ formula in
Sec.~\ref{sec:relative_determinant_TGTG}.

The same boundary-space reduction also identifies the
background-dressed one-plate transition operator. On the
one-dimensional boundary space associated with plate $i$, define
\begin{equation*}
	\mathcal T_{i,\partial}(\xi)
	=
	\lambda_i
	\left(
		\mathbb I_{\mathbb C}
		+
		\lambda_i\Gamma_iR_\xi\Gamma_i^\dagger
	\right)^{-1}.
	\label{eq:appB_T_operator_def}
\end{equation*}
Since
\begin{equation*}
	\Gamma_iR_\xi\Gamma_i^\dagger
	=
	G_{ii}(\xi),
\end{equation*}
this boundary-space operator is multiplication by the scalar
\begin{equation*}
	\tau_i(\xi)
	=
	\frac{\lambda_i}{1+\lambda_iG_{ii}(\xi)} .
	\label{eq:appB_tau_i}
\end{equation*}
In the customary distributional notation one may write formally
\begin{equation*}
	T_i(\xi)
	=
	\tau_i(\xi)\,|a_i\rangle\langle a_i| .
	\label{eq:appB_T_rank_one}
\end{equation*}
This last expression is shorthand for the corresponding
one-dimensional boundary-space transition operator; it does not define
a bounded rank-one operator on $L^2(\mathbb R)$. In the real-momentum
scattering problem, $\tau_i(\xi)$ is the analytic continuation of the
one-plate transition amplitude, up to the standard normalization factors
relating $T$-matrix elements to reflection and transmission
coefficients.

Consequently,
\begin{equation*}
	\Delta(\xi)
	=
	1-\tau_1(\xi)\tau_2(\xi)G_{12}G_{21}.
	\label{eq:appB_TGTG_scalar}
\end{equation*}
This is the scalar boundary-space form corresponding to the customary
formal notation $\det(1-T_1GT_2G)$.

Finally, the preceding argument is not special to two delta
interactions. More generally, point-supported perturbations admitting a
finite-dimensional boundary-space formulation lead to a
finite-dimensional Birman--Schwinger operator. The corresponding
relative determinant is therefore an ordinary determinant on that
boundary space, with entries constructed from the background Green
function and, when required by the boundary data, its derivatives
evaluated at the support points.

\section{Infrared expansion of the relative determinant}
\label{app:infrared_relative_determinant}

In this appendix we determine the infrared behaviour of the exact
relative determinant. Using the notation of Sec.~\ref{sec:relative_determinant_TGTG},
\begin{equation*}
	G_{ij}(\xi)
	=
	\frac{\psi_i\psi_j}{\xi^2}
	+
	g^\perp_{ij}(\xi),
	\qquad
	\psi_i=\psi_0(a_i),
\end{equation*}
where $g^\perp_{ij}(\xi)$ contains the contribution of the massive
fluctuations. Since the massive spectrum is separated from the zero mode
by a gap,
\begin{equation}
	g^\perp_{ij}(\xi)
	=
	g^\perp_{ij}(0)
	+
	O(\xi^2),
	\qquad
	\xi\to0^+ .
	\label{eq:appC_gperp_IR}
\end{equation}

For $\lambda_1,\lambda_2>0$, inserting this expansion into the exact
two-plate determinant gives
\begin{align}
	D_{12}(\xi)
	={}&
	\frac{1}{\xi^2}
	\Big[
	\lambda_1\psi_1^2
	+
	\lambda_2\psi_2^2
	\nonumber\\
	&\quad
	+
	\lambda_1\lambda_2
	\big(
	\psi_1^2 g^\perp_{22}(0)
	+
	\psi_2^2 g^\perp_{11}(0)
	\nonumber\\
	&\qquad\qquad
	-
	2\psi_1\psi_2 g^\perp_{12}(0)
	\big)
	\Big]
	+
	O(1).
	\label{eq:appC_D12_IR}
\end{align}
The leading $\xi^{-4}$ terms cancel identically because
$\psi_1^2\psi_2^2-(\psi_1\psi_2)^2=0$. On the other hand,
\begin{equation}
	D_1(\xi)D_2(\xi)
	=
	\frac{
		\lambda_1\lambda_2\psi_1^2\psi_2^2
	}{
		\xi^4
	}
	+
	O(\xi^{-2}) .
	\label{eq:appC_D1D2_IR}
\end{equation}
Therefore
\begin{equation}
	\Delta(\xi)
	=
	C_{\rm IR}\,\xi^2
	+
	O(\xi^4),
	\qquad
	\xi\to0^+,
	\label{eq:appC_Delta_IR_asymptotic}
\end{equation}
with
\begin{align}
	C_{\rm IR}
	={}&
	\frac{1}{
		\lambda_1\lambda_2\psi_1^2\psi_2^2
	}
	\Big[
	\lambda_1\psi_1^2
	+
	\lambda_2\psi_2^2
	\nonumber\\
	&\quad
	+
	\lambda_1\lambda_2
	\big(
	\psi_1^2 g^\perp_{22}(0)
	+
	\psi_2^2 g^\perp_{11}(0)
	\nonumber\\
	&\qquad\qquad
	-
	2\psi_1\psi_2 g^\perp_{12}(0)
	\big)
	\Big].
	\label{eq:appC_CIR_def}
\end{align}

The combination in parentheses is non-negative because it is the
quadratic form of the positive-semidefinite matrix with entries
$g^\perp_{ij}(0)$ on $(\psi_2,-\psi_1)$. Since
$\psi_i\neq0$ for finite $a_i$, Eq.~\eqref{eq:appC_CIR_def} therefore
gives
\begin{equation*}
	C_{\rm IR}>0,
	\qquad
	\lambda_1,\lambda_2>0 .
\end{equation*}

It follows that
\begin{equation}
	\log\Delta(\xi)
	=
	2\log\xi
	+
	\log C_{\rm IR}
	+
	O(\xi^2),
	\qquad
	\xi\to0^+ .
	\label{eq:appC_logDelta_IR}
\end{equation}
Consequently,
\begin{equation*}
	\int_0^\varepsilon
	d\xi\,
	\left|\log\Delta(\xi)\right|
	<
	\infty
\end{equation*}
for sufficiently small $\varepsilon>0$. Thus the zero-mode pole of the
P\"oschl--Teller resolvent produces only an integrable logarithmic
singularity in the interaction-energy integrand.

For $\lambda_i\geq0$ and $\xi>0$, the one- and two-plate
Birman--Schwinger determinants are positive. Hence $\Delta(\xi)>0$ and
$\log\Delta(\xi)$ has no branch ambiguity along the Euclidean
integration contour. If either coupling vanishes, the corresponding
plate is absent and $\Delta(\xi)=1$ identically.

\section{Birman--Schwinger equations for delta perturbations}
\label{app:birman_schwinger}

In this appendix we collect the Birman--Schwinger conditions needed in
the stability discussion of Sec.~\ref{sec:bound_states_stability}. Their
use for one-dimensional point interactions is standard
\cite{ReedSimonIV1978,SimonTraceIdeals2005,albeverio2005}.

Consider first a single plate with coupling $\lambda$ at $x=a$. A
negative normal mode may be written as
\begin{equation*}
	\omega^2=-\eta^2,
	\qquad
	\eta>0 .
\end{equation*}
The corresponding Birman--Schwinger condition is
\begin{equation}
	1+\lambda G_\eta^{\rm PT}(a,a)=0 .
	\label{eq:appD_single_BS_condition}
\end{equation}
For $\eta>0$, $G_\eta^{\rm PT}(a,a)>0$. Hence
Eq.~\eqref{eq:appD_single_BS_condition} has no solution for
$\lambda\geq0$. For $\lambda<0$, the instability of the unshifted
one-plate problem already follows directly from the zero-mode
quadratic-form argument given in Sec.~\ref{sec:bound_states_stability};
Eq.~\eqref{eq:appD_single_BS_condition} determines the corresponding
negative eigenvalue.

For two plates, the same Birman--Schwinger reduction gives the condition
\begin{equation}
	\det_{\mathbb C^2}
	\left[
	\mathbb I_2+\Lambda\mathbb G(\eta)
	\right]
	=
	0
	\label{eq:appD_two_BS_condition}
\end{equation}
for the existence of a negative normal mode. Here $\Lambda$ and
$\mathbb G$ are the coupling and Green-function matrices introduced in
Sec.~\ref{sec:finite_rank_plates}. In the manifestly stable sector
$\lambda_1,\lambda_2\geq0$, the quadratic form is non-negative, so
Eq.~\eqref{eq:appD_two_BS_condition} has no solution with $\eta>0$.

For the relative interaction energy, stability of the combined
two-plate operator is not sufficient by itself. The subtraction
prescription also contains the two isolated one-plate configurations,
and these must define stable fluctuation QFTs as well. Since any isolated
attractive plate destabilizes the zero mode in the unshifted
P\"oschl--Teller background, the relative construction used in the main
text is restricted to
\begin{equation*}
	\lambda_1\geq0,
	\qquad
	\lambda_2\geq0 .
\end{equation*}

Attractive plates can be accommodated after lifting the zero mode by a
positive mass shift,
\begin{equation*}
	K_{\rm PT}^{(\mu)}
	=
	K_{\rm PT}+\mu^2,
	\qquad
	\mu>0 .
\end{equation*}
No new Green function is required: the resolvent entering the
Birman--Schwinger equations is simply
$G_{\sqrt{\mu^2+\eta^2}}^{\rm PT}$. For a single plate, the threshold
at which a zero mode appears is therefore
\begin{equation*}
	1+\lambda G_\mu^{\rm PT}(a,a)=0 ,
\end{equation*}
or equivalently
\begin{equation}
	\lambda_c(a;\mu)
	=
	-\frac{1}{G_\mu^{\rm PT}(a,a)} .
	\label{eq:appD_lambda_critical}
\end{equation}
The shifted one-plate operator is non-negative for
$\lambda\geq\lambda_c(a;\mu)$. The corresponding two-plate stability
condition is obtained from Eq.~\eqref{eq:appD_two_BS_condition} by the
replacement
\begin{equation*}
	\mathbb G(\eta)
	\longrightarrow
	\mathbb G\!\left(\sqrt{\mu^2+\eta^2}\right),
\end{equation*}
together with the one-plate conditions required by the relative
subtraction.

In the limit $\mu\to0^+$, the zero-mode pole implies
$G_\mu^{\rm PT}(a,a)\to+\infty$, and therefore
\begin{equation*}
	\lambda_c(a;\mu)\to0^- .
\end{equation*}
Thus the unshifted limit reproduces the result used throughout the main
text: an isolated attractive delta plate is unstable.

\section{One-dimensional scattering amplitudes and Euclidean continuation}
\label{app:one_dimensional_scattering}

In this appendix we derive the scattering identity used in
Sec.~\ref{sec:scattering_interpretation}. We first work at real
momentum, where the quantities involved are ordinary one-dimensional
scattering amplitudes, and only afterwards continue them to the
imaginary-momentum axis relevant for the Euclidean vacuum-energy
formula. We use standard one-dimensional scattering conventions
\cite{Boya2008,Barlette2001,SanchezSoto2012} and the usual matching
conditions for point interactions \cite{albeverio2005}.

For continuum modes of the Pöschl--Teller operator,
$E(k)=1+k^2$ with $k>0$, and
\begin{equation*}
	\left[
	-\frac{d^2}{dx^2}
	-
	2\,\operatorname{sech}^2x
	\right]\psi_k(x)
	=
	k^2\psi_k(x).
\end{equation*}
A convenient pair of reflectionless background scattering solutions is
\begin{equation*}
	\phi_k^\pm(x)
	=
	e^{\pm ikx}
	\frac{\pm\tanh x-ik}{1-ik},
\end{equation*}
normalized by
\begin{equation*}
	\phi_k^\pm(x)\sim e^{\pm ikx},
	\qquad
	x\to\pm\infty .
\end{equation*}
With the convention $W[f,g]=fg'-f'g$, their Wronskian is
\begin{equation}
	W(k)
	=
	W[\phi_k^-,\phi_k^+]
	=
	-\frac{2ik(1+k^2)}{(1-ik)^2}.
	\label{eq:appE_real_channel_wronskian}
\end{equation}

Now place a delta plate of coupling $\lambda$ at $x=a$. Continuity of
the mode and the jump condition
\begin{equation*}
	\psi'(a^+)-\psi'(a^-)=\lambda\psi(a)
\end{equation*}
give the reflection amplitudes
\begin{align}
	r^R_{\lambda,a}(k)
	&=
	\frac{
		\lambda\,\phi_k^-(a)^2
	}{
		W(k)
		-
		\lambda\,\phi_k^-(a)\phi_k^+(a)
	},
	\label{eq:appE_real_right_reflection}
	\\
	r^L_{\lambda,a}(k)
	&=
	\frac{
		\lambda\,\phi_k^+(a)^2
	}{
		W(k)
		-
		\lambda\,\phi_k^-(a)\phi_k^+(a)
	}.
	\label{eq:appE_real_left_reflection}
\end{align}
The superscripts indicate the side from which the plate is seen by the
incoming wave. For two plates with $a_1<a_2$, the scattering factor
relevant to the interaction is the product of the reflection amplitude
of the left plate seen from the right and that of the right plate seen
from the left.

The reflection factors entering the vacuum-energy formula are obtained
by analytic continuation of these physical amplitudes,
\begin{equation*}
	k\longmapsto i\kappa,
	\qquad
	\kappa=\sqrt{1+\xi^2}.
\end{equation*}
The scattering solutions then become proportional to the decaying
Euclidean solutions of App.~\ref{app:PT_green_function},
\begin{equation*}
	\phi_{i\kappa}^{\pm}(x)
	=
	\frac{1}{1+\kappa}\,
	f_\xi^\pm(x),
\end{equation*}
and the corresponding Wronskians satisfy
\begin{equation}
	W(i\kappa)
	=
	\frac{W_\xi}{(1+\kappa)^2},
	\qquad
	W_\xi=-2\kappa(\kappa^2-1).
	\label{eq:appE_wronskian_continuation}
\end{equation}
Equations~\eqref{eq:appE_real_right_reflection} and
\eqref{eq:appE_real_left_reflection} therefore define the analytically
continued reflection factors
$\rho^R_{\lambda,a}(\xi)$ and $\rho^L_{\lambda,a}(\xi)$ used in
Sec.~\ref{sec:scattering_interpretation}.

To connect them with the relative determinant, recall from
App.~\ref{app:PT_green_function} that, for $a_1<a_2$,
\begin{align*}
	G_\xi^{\rm PT}(a_1,a_2)
	&=
	-\frac{
		f_\xi^-(a_1)f_\xi^+(a_2)
	}{
		W_\xi
	},
	\\
	G_\xi^{\rm PT}(a_i,a_i)
	&=
	-\frac{
		f_\xi^-(a_i)f_\xi^+(a_i)
	}{
		W_\xi
	},
	\qquad i=1,2 .
\end{align*}
Hence
\begin{equation*}
	1+\lambda_iG_{ii}(\xi)
	=
	\frac{
		W_\xi
		-
		\lambda_i f_\xi^-(a_i)f_\xi^+(a_i)
	}{
		W_\xi
	}.
\end{equation*}
Using these relations in the analytically continued reflection
amplitudes gives
\begin{equation}
\begin{aligned}
	\rho^R_{\lambda_1,a_1}(\xi)\,
	&\rho^L_{\lambda_2,a_2}(\xi)
	\\
	&=
	\frac{
		\lambda_1\lambda_2G_{12}(\xi)^2
	}{
		[1+\lambda_1G_{11}(\xi)]
		[1+\lambda_2G_{22}(\xi)]
	}.
\end{aligned}
\label{eq:appE_reflection_product_identity}
\end{equation}
This is precisely the two-plate scattering factor
$\mathcal R(\xi)$ defined in Eq.~\eqref{eq:secV_R_explicit}. Consequently,
\begin{equation}
	\Delta(\xi)
	=
	1-
	\rho^R_{\lambda_1,a_1}(\xi)\,
	\rho^L_{\lambda_2,a_2}(\xi),
	\label{eq:appE_Delta_reflection}
\end{equation}
which establishes the scattering interpretation used in
Sec.~\ref{sec:scattering_interpretation}. The amplitudes retain their
dependence on the individual plate positions because propagation and
one-plate scattering are both defined with respect to the fixed
Pöschl--Teller background.

As a consistency check, in the homogeneous massive limit one obtains
\begin{equation*}
\begin{aligned}
	\rho^R_{\lambda_1,a_1}(\xi)\,
	&\rho^L_{\lambda_2,a_2}(\xi)
	\\
	&=
	\frac{
		\lambda_1\lambda_2
		e^{-2\kappa(a_2-a_1)}
	}{
		(2\kappa+\lambda_1)
		(2\kappa+\lambda_2)
	},
\end{aligned}
\end{equation*}
which is the standard free massive double-delta result.

\begin{acknowledgments}

J.M.C. and G.S.G. acknowledge partial financial support from Grant
PID2021-123251NB-I00 funded by MCIN/AEI/10.13039/501100011033 and by
the European Union, and from the Department of Education of the Junta
de Castilla y León and FEDER Funds (Reference
No.~CLU-2025-1-02-IMUVA). G.S.G. also acknowledges financial support
from a University of Valladolid predoctoral contract cofunded by Banco
Santander (Grant No.~CONTPR-2024-310) and from the Formación de
Profesorado Universitario (FPU) program of the Spanish Ministry of
Science, Innovation and Universities (Grant No.~FPU24/01715).
I.C.P. acknowledges partial financial support from Spanish MINECO/FEDER
Grant PID2024-160228NB-I00, funded by MCIN/AEI/10.13039/501100011033
and the European Regional Development Fund (ERDF, ``A way of making
Europe''), and by Grant E21-23R funded by the Government of Aragon and
the European Union.

J.M.C. and I.C.P. wish to express their deepest gratitude to Juan Mateos
Guilarte, whose teaching, generosity, and guidance have shaped their
scientific formation over many years. For J.M.C. in particular, his
influence reaches far beyond academia: he has been a constant source of
wisdom, encouragement, and guidance since long before the beginning of
his scientific career. The authors thank Jesús Clemente Gallardo, Manuel Donaire, and Beatriz Gómez López for valuable discussions.

\end{acknowledgments}

%

\begin{thebibliography}{21}%
\makeatletter
\providecommand \@ifxundefined [1]{%
 \@ifx{#1\undefined}
}%
\providecommand \@ifnum [1]{%
 \ifnum #1\expandafter \@firstoftwo
 \else \expandafter \@secondoftwo
 \fi
}%
\providecommand \@ifx [1]{%
 \ifx #1\expandafter \@firstoftwo
 \else \expandafter \@secondoftwo
 \fi
}%
\providecommand \natexlab [1]{#1}%
\providecommand \enquote  [1]{``#1''}%
\providecommand \bibnamefont  [1]{#1}%
\providecommand \bibfnamefont [1]{#1}%
\providecommand \citenamefont [1]{#1}%
\providecommand \href@noop [0]{\@secondoftwo}%
\providecommand \href [0]{\begingroup \@sanitize@url \@href}%
\providecommand \@href[1]{\@@startlink{#1}\@@href}%
\providecommand \@@href[1]{\endgroup#1\@@endlink}%
\providecommand \@sanitize@url [0]{\catcode `\\12\catcode `\$12\catcode
  `\&12\catcode `\#12\catcode `\^12\catcode `\_12\catcode `\%12\relax}%
\providecommand \@@startlink[1]{}%
\providecommand \@@endlink[0]{}%
\providecommand \url  [0]{\begingroup\@sanitize@url \@url }%
\providecommand \@url [1]{\endgroup\@href {#1}{\urlprefix }}%
\providecommand \urlprefix  [0]{URL }%
\providecommand \Eprint [0]{\href }%
\providecommand \doibase [0]{https://doi.org/}%
\providecommand \selectlanguage [0]{\@gobble}%
\providecommand \bibinfo  [0]{\@secondoftwo}%
\providecommand \bibfield  [0]{\@secondoftwo}%
\providecommand \translation [1]{[#1]}%
\providecommand \BibitemOpen [0]{}%
\providecommand \bibitemStop [0]{}%
\providecommand \bibitemNoStop [0]{.\EOS\space}%
\providecommand \EOS [0]{\spacefactor3000\relax}%
\providecommand \BibitemShut  [1]{\csname bibitem#1\endcsname}%
\let\auto@bib@innerbib\@empty
\bibitem [{\citenamefont {Casimir}(1948)}]{casimir1948}%
  \BibitemOpen
  \bibfield  {author} {\bibinfo {author} {\bibfnamefont {H.~B.~G.}\
  \bibnamefont {Casimir}},\ }\href@noop {} {\bibfield  {journal} {\bibinfo
  {journal} {Proceedings of the Koninklijke Nederlandse Akademie van
  Wetenschappen}\ }\textbf {\bibinfo {volume} {51}},\ \bibinfo {pages} {793}
  (\bibinfo {year} {1948})}\BibitemShut {NoStop}%
\bibitem [{\citenamefont {Milton}(2001)}]{milton2001book}%
  \BibitemOpen
  \bibfield  {author} {\bibinfo {author} {\bibfnamefont {K.~A.}\ \bibnamefont
  {Milton}},\ }\href@noop {} {\emph {\bibinfo {title} {The Casimir Effect:
  Physical Manifestations of Zero-Point Energy}}}\ (\bibinfo  {publisher}
  {World Scientific},\ \bibinfo {address} {Singapore},\ \bibinfo {year}
  {2001})\BibitemShut {NoStop}%
\bibitem [{\citenamefont {Bordag}\ \emph {et~al.}(2009)\citenamefont {Bordag},
  \citenamefont {Klimchitskaya}, \citenamefont {Mohideen},\ and\ \citenamefont
  {Mostepanenko}}]{bordag2009book}%
  \BibitemOpen
  \bibfield  {author} {\bibinfo {author} {\bibfnamefont {M.}~\bibnamefont
  {Bordag}}, \bibinfo {author} {\bibfnamefont {G.~L.}\ \bibnamefont
  {Klimchitskaya}}, \bibinfo {author} {\bibfnamefont {U.}~\bibnamefont
  {Mohideen}},\ and\ \bibinfo {author} {\bibfnamefont {V.~M.}\ \bibnamefont
  {Mostepanenko}},\ }\href@noop {} {\emph {\bibinfo {title} {Advances in the
  Casimir Effect}}}\ (\bibinfo  {publisher} {Oxford University Press},\
  \bibinfo {address} {Oxford},\ \bibinfo {year} {2009})\BibitemShut {NoStop}%
\bibitem [{\citenamefont {Bordag}\ \emph {et~al.}(1985)\citenamefont {Bordag},
  \citenamefont {Robaschik},\ and\ \citenamefont
  {Wieczorek}}]{bordag1992delta}%
  \BibitemOpen
  \bibfield  {author} {\bibinfo {author} {\bibfnamefont {M.}~\bibnamefont
  {Bordag}}, \bibinfo {author} {\bibfnamefont {D.}~\bibnamefont {Robaschik}},\
  and\ \bibinfo {author} {\bibfnamefont {E.}~\bibnamefont {Wieczorek}},\ }\href
  {https://doi.org/10.1016/0003-4916(85)90088-8} {\bibfield  {journal}
  {\bibinfo  {journal} {Annals of Physics}\ }\textbf {\bibinfo {volume}
  {165}},\ \bibinfo {pages} {192} (\bibinfo {year} {1985})}\BibitemShut
  {NoStop}%
\bibitem [{\citenamefont {Milton}(2004)}]{milton2004casimir}%
  \BibitemOpen
  \bibfield  {author} {\bibinfo {author} {\bibfnamefont {K.~A.}\ \bibnamefont
  {Milton}},\ }\href {https://doi.org/10.1088/0305-4470/37/38/R01}
  {\bibfield  {journal} {\bibinfo  {journal} {Journal of Physics A:
  Mathematical and General}\ }\textbf {\bibinfo {volume} {37}},\
  \bibinfo {pages} {R209} (\bibinfo {year} {2004})}\BibitemShut {NoStop}%
\bibitem [{\citenamefont {Parashar}\ \emph {et~al.}(2012)\citenamefont
  {Parashar}, \citenamefont {Milton}, \citenamefont {Shajesh},\ and\
  \citenamefont {Schaden}}]{parashar2012delta}%
  \BibitemOpen
  \bibfield  {author} {\bibinfo {author} {\bibfnamefont {P.}\ \bibnamefont
  {Parashar}}, \bibinfo {author} {\bibfnamefont {K.~A.}\ \bibnamefont
  {Milton}}, \bibinfo {author} {\bibfnamefont {K.~V.}\ \bibnamefont
  {Shajesh}},\ and\ \bibinfo {author} {\bibfnamefont {M.}\ \bibnamefont
  {Schaden}},\ }\href {https://doi.org/10.1103/PhysRevD.86.085021}
  {\bibfield  {journal} {\bibinfo  {journal} {Physical Review D}\ }\textbf
  {\bibinfo {volume} {86}},\ \bibinfo {pages} {085021} (\bibinfo {year}
  {2012})}\BibitemShut {NoStop}%
\bibitem [{\citenamefont {Muñoz-Castañeda}\ \emph {et~al.}(2013)\citenamefont
  {Muñoz-Castañeda}, \citenamefont {Mateos~Guilarte},\ and\ \citenamefont
  {Moreno~Mosquera}}]{munoz2013prd}%
  \BibitemOpen
  \bibfield  {author} {\bibinfo {author} {\bibfnamefont {J.~M.}\ \bibnamefont
  {Muñoz-Castañeda}}, \bibinfo {author} {\bibfnamefont {J.}~\bibnamefont
  {Mateos~Guilarte}},\ and\ \bibinfo {author} {\bibfnamefont {A.}~\bibnamefont
  {Moreno~Mosquera}},\ }\href {https://doi.org/10.1103/PhysRevD.87.105020}
  {\bibfield  {journal} {\bibinfo  {journal} {Physical Review D}\ }\textbf
  {\bibinfo {volume} {87}},\ \bibinfo {pages} {105020} (\bibinfo {year}
  {2013})},\ \Eprint {https://arxiv.org/abs/1305.2054} {arXiv:1305.2054
  [hep-th]} \BibitemShut {NoStop}%
\bibitem [{\citenamefont {Muñoz-Castañeda}\ and\ \citenamefont
  {Mateos~Guilarte}(2015)}]{munoz2015ddp}%
  \BibitemOpen
  \bibfield  {author} {\bibinfo {author} {\bibfnamefont {J.~M.}\ \bibnamefont
  {Muñoz-Castañeda}}\ and\ \bibinfo {author} {\bibfnamefont {J.}~\bibnamefont
  {Mateos~Guilarte}},\ }\href {https://doi.org/10.1103/PhysRevD.91.025028}
  {\bibfield  {journal} {\bibinfo  {journal} {Physical Review D}\ }\textbf
  {\bibinfo {volume} {91}},\ \bibinfo {pages} {025028} (\bibinfo {year}
  {2015})},\ \Eprint {https://arxiv.org/abs/1411.5773} {arXiv:1411.5773
  [hep-th]} \BibitemShut {NoStop}%
\bibitem [{\citenamefont {Vilenkin}\ and\ \citenamefont
  {Shellard}(1994)}]{vilenkinShellard1994}%
  \BibitemOpen
  \bibfield  {author} {\bibinfo {author} {\bibfnamefont {A.}~\bibnamefont
  {Vilenkin}}\ and\ \bibinfo {author} {\bibfnamefont {E.~P.~S.}\ \bibnamefont
  {Shellard}},\ }\href@noop {} {\emph {\bibinfo {title} {Cosmic Strings and
  Other Topological Defects}}}\ (\bibinfo  {publisher} {Cambridge University
  Press},\ \bibinfo {year} {1994})\BibitemShut {NoStop}%
\bibitem [{\citenamefont {Ipser}\ and\ \citenamefont
  {Sikivie}(1984)}]{ipserSikivie1984}%
  \BibitemOpen
  \bibfield  {author} {\bibinfo {author} {\bibfnamefont {J.}~\bibnamefont
  {Ipser}}\ and\ \bibinfo {author} {\bibfnamefont {P.}~\bibnamefont
  {Sikivie}},\ }\href {https://doi.org/10.1103/PhysRevD.30.712} {\bibfield
  {journal} {\bibinfo  {journal} {Physical Review D}\ }\textbf {\bibinfo
  {volume} {30}},\ \bibinfo {pages} {712} (\bibinfo {year} {1984})}\BibitemShut
  {NoStop}%
\bibitem [{\citenamefont {Rajaraman}(1982)}]{rajaraman1982}%
  \BibitemOpen
  \bibfield  {author} {\bibinfo {author} {\bibfnamefont {R.}~\bibnamefont
  {Rajaraman}},\ }\href@noop {} {\emph {\bibinfo {title} {Solitons and
  Instantons: An Introduction to Solitons and Instantons in Quantum Field
  Theory}}}\ (\bibinfo  {publisher} {North-Holland},\ \bibinfo {year}
  {1982})\BibitemShut {NoStop}%
\bibitem [{\citenamefont {Dashen}\ \emph {et~al.}(1974)\citenamefont {Dashen},
  \citenamefont {Hasslacher},\ and\ \citenamefont {Neveu}}]{DHN1974}%
  \BibitemOpen
  \bibfield  {author} {\bibinfo {author} {\bibfnamefont {R.~F.}\ \bibnamefont
  {Dashen}}, \bibinfo {author} {\bibfnamefont {B.}~\bibnamefont {Hasslacher}},\
  and\ \bibinfo {author} {\bibfnamefont {A.}~\bibnamefont {Neveu}},\ }\href
  {https://doi.org/10.1103/PhysRevD.10.4130} {\bibfield  {journal} {\bibinfo
  {journal} {Physical Review D}\ }\textbf {\bibinfo {volume} {10}},\ \bibinfo
  {pages} {4130} (\bibinfo {year} {1974})}\BibitemShut {NoStop}%
\bibitem [{\citenamefont {Reed}\ and\ \citenamefont
  {Simon}(1978)}]{ReedSimonIV1978}%
  \BibitemOpen
  \bibfield  {author} {\bibinfo {author} {\bibfnamefont {M.}~\bibnamefont
  {Reed}}\ and\ \bibinfo {author} {\bibfnamefont {B.}~\bibnamefont {Simon}},\
  }\href@noop {} {\emph {\bibinfo {title} {Methods of Modern Mathematical
  Physics IV: Analysis of Operators}}}\ (\bibinfo  {publisher} {Academic
  Press},\ \bibinfo {address} {New York},\ \bibinfo {year} {1978})\BibitemShut
  {NoStop}%
\bibitem [{\citenamefont {Simon}(2005)}]{SimonTraceIdeals2005}%
  \BibitemOpen
  \bibfield  {author} {\bibinfo {author} {\bibfnamefont {B.}~\bibnamefont
  {Simon}},\ }\href {https://doi.org/10.1090/surv/120} {\emph {\bibinfo {title}
  {Trace Ideals and Their Applications}}},\ \bibinfo {edition} {2nd}\ ed.,\
  \bibinfo {series} {Mathematical Surveys and Monographs}, Vol.\ \bibinfo
  {volume} {120}\ (\bibinfo  {publisher} {American Mathematical Society},\
  \bibinfo {address} {Providence, RI},\ \bibinfo {year} {2005})\BibitemShut
  {NoStop}%
\bibitem [{\citenamefont {Kenneth}\ and\ \citenamefont
  {Klich}(2008)}]{KennethKlich2008}%
  \BibitemOpen
  \bibfield  {author} {\bibinfo {author} {\bibfnamefont {O.}~\bibnamefont
  {Kenneth}}\ and\ \bibinfo {author} {\bibfnamefont {I.}~\bibnamefont
  {Klich}},\ }\href {https://doi.org/10.1103/PhysRevB.78.014103} {\bibfield
  {journal} {\bibinfo  {journal} {Physical Review B}\ }\textbf {\bibinfo
  {volume} {78}},\ \bibinfo {pages} {014103} (\bibinfo {year} {2008})},\
  \Eprint {https://arxiv.org/abs/0707.4017} {arXiv:0707.4017 [quant-ph]}
  \BibitemShut {NoStop}%
\bibitem [{\citenamefont {Santamaría-Sanz}(2024)}]{casimirTransferSG}%
  \BibitemOpen
  \bibfield  {author} {\bibinfo {author} {\bibfnamefont {L.}~\bibnamefont
  {Santamaría-Sanz}},\ }\href {https://doi.org/10.1093/ptep/ptae059}
  {\bibfield  {journal} {\bibinfo  {journal} {Progress of Theoretical and
  Experimental Physics}\ }\textbf {\bibinfo {volume} {2024}},\ \bibinfo {pages}
  {053A03} (\bibinfo {year} {2024})},\ \Eprint
  {https://arxiv.org/abs/2305.01438} {arXiv:2305.01438 [hep-th]} \BibitemShut
  {NoStop}%
\bibitem [{\citenamefont {Albeverio}\ \emph {et~al.}(2005)\citenamefont
  {Albeverio}, \citenamefont {Gesztesy}, \citenamefont {H{\o}egh-Krohn},\ and\
  \citenamefont {Holden}}]{albeverio2005}%
  \BibitemOpen
  \bibfield  {author} {\bibinfo {author} {\bibfnamefont {S.}~\bibnamefont
  {Albeverio}}, \bibinfo {author} {\bibfnamefont {F.}~\bibnamefont {Gesztesy}},
  \bibinfo {author} {\bibfnamefont {R.}~\bibnamefont {H{\o}egh-Krohn}},\ and\
  \bibinfo {author} {\bibfnamefont {H.}~\bibnamefont {Holden}},\ }\href
  {https://doi.org/10.1090/chel/350} {\emph {\bibinfo {title} {Solvable Models
  in Quantum Mechanics}}},\ \bibinfo {edition} {2nd}\ ed.,\ \bibinfo {series}
  {AMS Chelsea Publishing}, Vol.\ \bibinfo {volume} {350}\ (\bibinfo
  {publisher} {American Mathematical Society},\ \bibinfo {address} {Providence,
  RI},\ \bibinfo {year} {2005})\BibitemShut {NoStop}%
\bibitem [{\citenamefont {Barton}\ and\ \citenamefont
  {Waxman}(1994)}]{barton1993waxman}%
  \BibitemOpen
  \bibfield  {author} {\bibinfo {author} {\bibfnamefont {G.}~\bibnamefont
  {Barton}}\ and\ \bibinfo {author} {\bibfnamefont {D.}~\bibnamefont
  {Waxman}},\ }\href {https://doi.org/10.1088/0305-4470/27/10/026} {\bibfield
  {journal} {\bibinfo  {journal} {Journal of Physics A: Mathematical and
  General}\ }\textbf {\bibinfo {volume} {27}},\ \bibinfo {pages} {3311}
  (\bibinfo {year} {1994})}\BibitemShut {NoStop}%
\bibitem [{\citenamefont {Asorey}\ and\ \citenamefont
  {Mu{\~n}oz-Casta{\~n}eda}(2013)}]{MunozCastaneda2013}%
  \BibitemOpen
  \bibfield  {author} {\bibinfo {author} {\bibfnamefont {M.}~\bibnamefont
  {Asorey}}\ and\ \bibinfo {author} {\bibfnamefont {J.~M.}\ \bibnamefont
  {Mu{\~n}oz-Casta{\~n}eda}},\ }\href
  {https://doi.org/10.1016/j.nuclphysb.2013.06.014}
  {\bibfield  {journal} {\bibinfo  {journal} {Nucl. Phys. B}\ }\textbf
  {\bibinfo {volume} {874}},\ \bibinfo {pages} {852} (\bibinfo {year}
  {2013})}\BibitemShut {NoStop}%
\bibitem [{\citenamefont {Mu{\~n}oz-Casta{\~n}eda}\ \emph
  {et~al.}(2026)\citenamefont {Mu{\~n}oz-Casta{\~n}eda}, \citenamefont
  {Cavero-Pel{\'a}ez},\ and\ \citenamefont
  {Sancho-Garrido}}]{MunozCastanedaTopologicalCasimir_inprep}%
  \BibitemOpen
  \bibfield  {author} {\bibinfo {author} {\bibfnamefont {J.~M.}\ \bibnamefont
  {Mu{\~n}oz-Casta{\~n}eda}}, \bibinfo {author} {\bibfnamefont
  {I.}~\bibnamefont {Cavero-Pel{\'a}ez}},\ and\ \bibinfo {author}
  {\bibfnamefont {G.}~\bibnamefont {Sancho-Garrido}},\ }\href@noop {} {\bibinfo
  {title} {Localized zero modes enhance massive Casimir interactions}}
  (\bibinfo {year} {2026}),\ \Eprint {https://arxiv.org/abs/2607.05530}
  {arXiv:2607.05530 [hep-th]} \BibitemShut {NoStop}%
\bibitem [{\citenamefont {Boya}(2008)}]{Boya2008}%
  \BibitemOpen
  \bibfield  {author} {\bibinfo {author} {\bibfnamefont {L.~J.}\ \bibnamefont
  {Boya}},\ }\href {https://doi.org/10.1393/ncr/i2008-10030-4} {\bibfield
  {journal} {\bibinfo  {journal} {La Rivista del Nuovo Cimento}\ }\textbf
  {\bibinfo {volume} {31}},\ \bibinfo {pages} {75} (\bibinfo {year}
  {2008})}\BibitemShut {NoStop}%
\bibitem [{\citenamefont {Barlette}\ \emph {et~al.}(2001)\citenamefont
  {Barlette}, \citenamefont {Leite},\ and\ \citenamefont
  {Adhikari}}]{Barlette2001}%
  \BibitemOpen
  \bibfield  {author} {\bibinfo {author} {\bibfnamefont {V.~E.}\ \bibnamefont
  {Barlette}}, \bibinfo {author} {\bibfnamefont {M.~M.}\ \bibnamefont
  {Leite}},\ and\ \bibinfo {author} {\bibfnamefont {S.~K.}\ \bibnamefont
  {Adhikari}},\ }\href {https://doi.org/10.1119/1.1371015} {\bibfield
  {journal} {\bibinfo  {journal} {American Journal of Physics}\ }\textbf
  {\bibinfo {volume} {69}},\ \bibinfo {pages} {1010} (\bibinfo {year}
  {2001})},\ \Eprint {https://arxiv.org/abs/quant-ph/0012087}
  {arXiv:quant-ph/0012087} \BibitemShut {NoStop}%
\bibitem [{\citenamefont {S{\'a}nchez-Soto}\ \emph {et~al.}(2012)\citenamefont
  {S{\'a}nchez-Soto}, \citenamefont {Monz{\'o}n}, \citenamefont {Barriuso},\
  and\ \citenamefont {Cari{\~n}ena}}]{SanchezSoto2012}%
  \BibitemOpen
  \bibfield  {author} {\bibinfo {author} {\bibfnamefont {L.~L.}\ \bibnamefont
  {S{\'a}nchez-Soto}}, \bibinfo {author} {\bibfnamefont {J.~J.}\ \bibnamefont
  {Monz{\'o}n}}, \bibinfo {author} {\bibfnamefont {A.~G.}\ \bibnamefont
  {Barriuso}},\ and\ \bibinfo {author} {\bibfnamefont {J.~F.}\ \bibnamefont
  {Cari{\~n}ena}},\ }\href {https://doi.org/10.1016/j.physrep.2011.10.002}
  {\bibfield  {journal} {\bibinfo  {journal} {Physics Reports}\ }\textbf
  {\bibinfo {volume} {513}},\ \bibinfo {pages} {191} (\bibinfo {year}
  {2012})}\BibitemShut {NoStop}%
\end{thebibliography}

%

%
\end{document}